%% file: arxiv_resub.tex
\documentclass[aps,prl,reprint]{revtex4-1}
\usepackage{amsmath,amssymb}
\usepackage{pifont}
\usepackage[font=small]{caption}   
\usepackage{subcaption}
\usepackage[noline,ruled]{algorithm2e}
\usepackage{algpseudocode}
\usepackage{xcolor}
\usepackage{graphicx}
\usepackage{booktabs}
\usepackage{dsfont}
\usepackage{bbm}
\usepackage{hyperref}
\usepackage{url}
\usepackage{bbm}
\hypersetup{
    colorlinks=true,
    linkcolor=red,
    citecolor=blue,
    filecolor=magenta,      
    urlcolor=red,
    pdftitle={Overleaf Example},
    pdfpagemode=FullScreen,
    }
\SetKwInOut{Input}{Require}
\SetKwInOut{Output}{Ensure}
\SetInd{2em}{2em} 

\newcommand{\sign}{\mathrm{sgn}}

\newcommand{\cP}{\mathcal{P}}

\DeclareUnicodeCharacter{00A0}{~}

\renewcommand{\arraystretch}{1.3}
\begin{document}

\title{Dynamical Learning in Deep Asymmetric Recurrent Neural Networks}

\author{Davide Badalotti}
\altaffiliation{Alphabetical by last name.\\}
\author{Carlo Baldassi}
\altaffiliation{Alphabetical by last name.\\}
\author{Marc Mézard}
\altaffiliation{Alphabetical by last name.\\}
\author{Mattia Scardecchia}
\altaffiliation{Alphabetical by last name.\\}
\author{Riccardo Zecchina}
\altaffiliation{Alphabetical by last name.\\}
\affiliation{Department of Computing Sciences, Bocconi University, Milan, Italy}

\begin{abstract}
We investigate recurrent neural networks with asymmetric interactions and demonstrate that the inclusion of self-couplings or sparse excitatory inter-module connections leads to the emergence of a densely connected manifold of dynamically accessible stable configurations. This representation manifold is exponentially large in system size and is reachable through simple local dynamics, despite constituting a subdominant subset of the global configuration space. We further show that learning can be implemented directly on this structure via a fully local, gradient-free mechanism that selectively stabilizes a single task-relevant network configuration. Unlike error-driven or contrastive learning schemes, this approach does not require explicit comparisons between network states obtained with and without output supervision. Instead, transient supervisory signals bias the dynamics toward the representation manifold, after which local plasticity consolidates the attained configuration, effectively shaping the latent representation space. Numerical evaluations on standard image classification benchmarks indicate performance comparable to that of multilayer perceptrons trained using backpropagation. More generally, these results suggest that the dynamical accessibility of fixed points and the stabilization of internal network dynamics constitute viable alternative principles for learning in recurrent systems, with conceptual links to statistical physics and potential implications for biologically motivated and neuromorphic computing architectures.
\end{abstract}

\maketitle

The remarkable success of contemporary artificial intelligence systems, driven by large-scale artificial neural networks (ANNs), has stimulated intense research into their underlying computational principles. Currently, the dominant training paradigm for ANNs is empirical risk minimization via backpropagation \cite{lecun2015_nature_deep}. While effective, this approach requires massive datasets and computational resources, and is generally considered incompatible with the goal of explaining computation in biological neural circuits \cite{neuroAI_panel_2024,ororbia2024review}.
Furthermore, the architectures underlying today’s state-of-the-art models \cite{lecun2015_nature_deep, sohl2015_diffusion_deep, vaswani2017_attention} fail to leverage the rich dynamical behaviors inherent to biological neural networks. A major scientific challenge thus lies in the design of ANNs that learn and operate more like biological brains, without relying on global gradient computation.

To this end, with feed-forward architectures, some approaches modify the backward pass by using learned inverses \cite{Lee_target_prop_2014} or random projections \cite{Lillicrap_feedback_alignment_2016} to propagate the error signal. Others replace it, either with a dynamics that updates activations and weights to make the forward pass internally consistent while forcing a desired output \cite{Rao_predictive_coding_1999, millidge2022predictivecodingtheoreticalexperimental}, or with an additional forward pass on synthetic data that must be distinguished from real data through a learned local criterion \cite{Hinton_forward_forward_2022}.
With recurrent architectures, which is the setting of the present work, Reservoir Computing \cite{nakajima_reservoir_2021} sidesteps the credit assignment problem by leveraging a fixed recurrent module, while spiking neural networks propose more realistic neuron models but are typically trained via Backpropagation Through Time \cite{lee_spiking_training_2016}.
The recently proposed Neural Langevin Machine \cite{yu2025neural} stabilizes the chaotic dynamics by introducing an Onsager-type feedback term, and applies a local contrastive asymmetric learning rule to train a generative model.
Finally, Equilibrium Propagation perturbs the dynamics of an energy-based model and uses a contrastive Hebbian rule to estimate the gradient of a global loss function \cite{Scellier_equilibrium_prop_2017}.

In this work we introduce an optimization-free learning paradigm based on stabilizing fixed points of the dynamics in asymmetric recurrent neural networks. Learning is fully distributed, relies only on local information, and does not require symmetry or a distinction between training and inference. Crucially, the learning step operates on a single task-relevant configuration: unlike contrastive schemes, it does not compare states obtained with and without output supervision, but directly stabilizes the reached configuration to shape the latent space. Using standard benchmarks, we show that this approach achieves performance comparable to a multilayer perceptron trained with backpropagation with the same number of parameters, highlighting its relevance for artificial intelligence.

Our model is based on a \emph{core module} which is a network of binary neurons interacting with asymmetric couplings. While these models have been studied for a long time in the context of computational neuroscience, most studies have focused on their chaotic regime \cite{sompolinsky1988chaos,stern2014dynamics}, or on the edge of chaos, for signal processing applications. At odds with these approaches, we introduce an excitatory self-coupling in the core module, or alternatively sparse and strong excitatory couplings between several core modules, and we show that with well-tuned strength of the excitation this leads to the appearance of an accessible, dense, connected cluster of stable fixed points 
--  a high local entropy region  \cite{baldassi_prl_2015} -- which we call the 
internal Representation Manifold (RM).

For simplicity we adopt a discrete-time, binary formulation. Our results concern the structure of stationary states, for which the discrete–continuous distinction is not essential, and preliminary simulations indicate that the framework extends to continuous time. Binary activations allow us to isolate robust qualitative effects without additional technical complications.

The geometry of the RM can be exploited for learning by mapping input-output associations onto fixed points of the dynamics, as follows. Given an input and a desired output (label), the network starts in a neutral configuration and initially evolves under the influence of the external input. After a short time, a transient supervisory signal from the output appears, driving the dynamics into a fixed point $s'$. Then, the supervisory signal is removed and the network settles into a second configuration $s^*$. Finally, learning happens through a local synaptic plasticity rule that stabilizes $s^*$. As it turns out, in the RM phase, $s'$ and $s^*$ are close to each other and the synaptic reinforcement leads to a stabilization of the input-output association: exposed to the same input again, without any supervision, the network state will evolve into a configuration that is mapped into the correct output by a learned linear readout. This procedure implements a simple, fully local learning mechanism grounded in the dynamics of asymmetric deep recurrent networks, which does not use error gradient and does not require any additional feedback circuit.

The paper is organized as follows. First, we define the model and present the statistical mechanics analysis concerning the onset of the RM. Next, we define the learning process and present the results of extensive numerical experiments. 
Finally, we study the role of the RM geometry for learning.

\textbf{Core Module.} The core module is a network of binary neurons with state \(s \in \{-1, 1\}^N\) and an \(N \times N\) diluted asymmetric interaction
matrix \(J\). The off-diagonal matrix elements $J_{ij},i\neq j$ are zero with probability $1-\rho$, and with probability $\rho$ they are drawn independently from an unbiased Gaussian distribution of variance $\frac{1}{\rho N} $.
We introduce a self-interaction term \(J_{ii} = J_D \ge 0\) for all \(i\), which softens the fixed-point constraints and affects the dynamics. 
The network evolves in discrete time by aligning each neuron to its local field through the  update rule: 
\begin{equation}
s_i \leftarrow \sign\Big( \sum_{j \neq i} J_{ij} s_j +J_D s_i  \Big),
\label{eq: dyn_core}
\end{equation}
with $\sign$ being the sign function. Throughout this study we adopt a synchronous dynamical scheme for the updates.
Due to the asymmetry of the interactions, the dynamics is not governed by a Lyapunov function and the structure of fixed points is subtle.
While for $J_D=0$ the network is chaotic and no fixed-points exist, for $J_D>0$ there exist exponentially many fixed points, which however remain unreachable to the dynamics. Above a critical value of $J_D$, there appears a subdominant cluster of fixed points, the RM, which can be accessed by various types of algorithms including (for large enough $J_D$) the simple iterative dynamics (\ref{eq: dyn_core}).
This phenomenology is reminiscent of what was found in supervised learning with non-convex, feed-forward classifiers \cite{baldassi_prl_2015,baldassi_pnas_2016,baldassi2021unveiling,barbier2025finding}. 

When using such a network for learning, the influence of the input -- and possibly of the output label -- on the network is mediated by linear input and output layers, respectively. These influence the dynamics of the system and can be trained jointly with the internal couplings using the same local learning rule.

\begin{figure}[h!]
    \centering
    \includegraphics[width=1.0\linewidth]{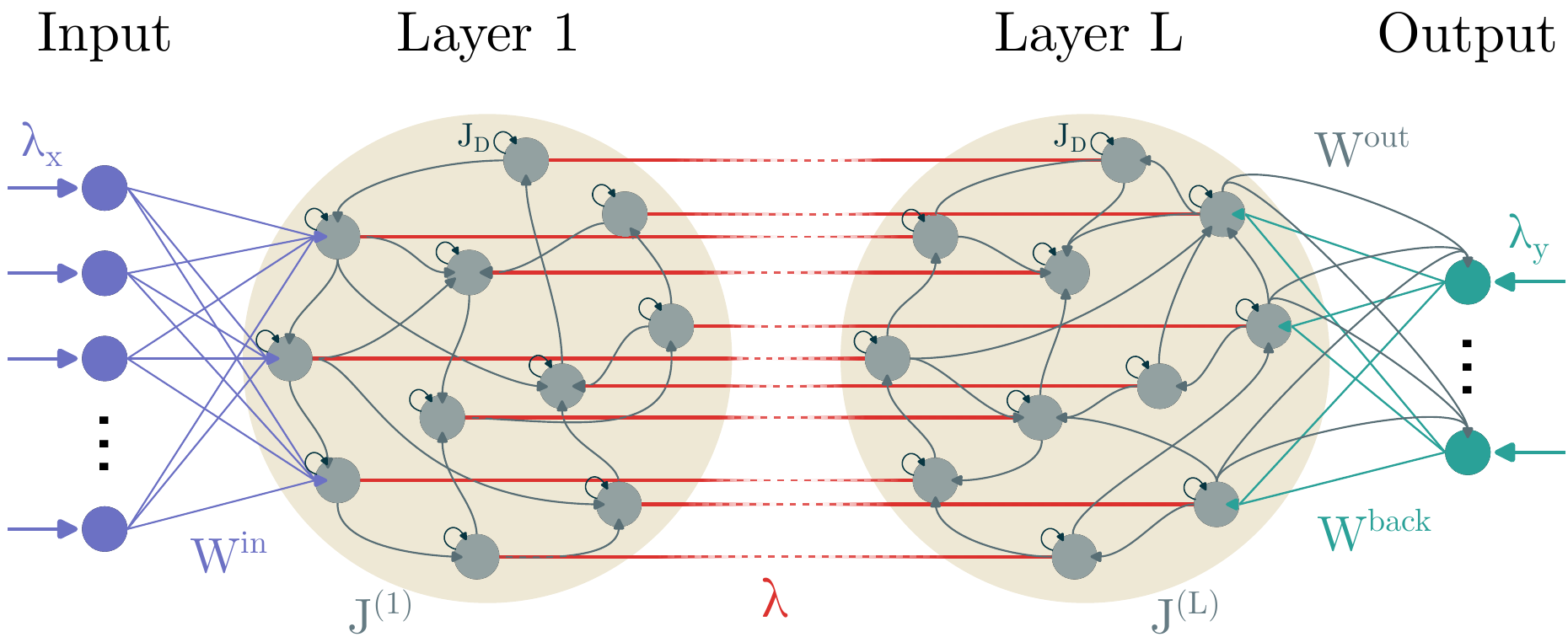}
    \caption{A multilayer chain model with sparse inter module excitatory couplings, and input/output layers.
    }
    \captionsetup{format=plain, font=small, labelfont=bf}
    \label{fig:models}
\end{figure}

\textbf{Multilayer Chain.} As a minimal model of hierarchical processing, we consider a chain of \(L\) core modules like the one above, with states
\(s^{(l)} \in \{-1, 1\}^{N}\),  and independent interaction matrices \(J^{(l)} \in \mathbb{R}^{N \times N}\), \(l = 1, \dots, L\). 
We introduce a positive interaction between homologous neurons in adjacent layers with uniform strength \(\lambda \ge 0\):
\begin{equation}
    s_i^{(l)} \leftarrow \sign\Big(  \sum_{j \neq i} J_{ij}^{(l)} s_j^{(l)}+ J_D s_i^{(l)} + \lambda (s_i^{(l-1)} + s_i^{(l+1)})\Big)  \, ,
    \label{eq:multilayer-chain-update}
\end{equation}
with the understanding that \(s_i^{(0)} \equiv s_i^{(L+1)} \equiv 0\). 
This reduces to the core model for \(L=1\). As we shall see, the excitatory sparse interactions across modules governed by $\lambda$ play a very similar role to the diagonal coupling $J_D$; in particular, when $\lambda$ is large enough the dynamics converges to  the RM even when $J_D=0$. 
Fig. \ref{fig:models} provides a sketch of the chain.


\textbf{Analysis of the geometry of fixed points.} To analyze the geometric structure of the fixed points and their accessibility,  we first compute the typical number ${\mathcal N}_\mathrm{fp}$ of fixed points of the multilayer chain in absence of external inputs, before learning takes place. Since this number is exponentially large, we shall compute the associated entropy density $S_\mathrm{fp}=\lim_{N\to\infty} \frac{1}{L N} \log \mathcal{N}_\mathrm{fp}$.  
As $S_\mathrm{fp}$ is expected to self-average, we are interested in the quenched average $\frac{1}{L N} \mathbb{E} \log \mathcal{N}_\mathrm{fp}$, where the expectation is taken with respect to the distribution of off-diagonal couplings $J_{ij}$. It turns out that in the large $N$ limit the `annealed average', i.e. the easily-computed first-moment upper bound $\frac{1}{L N}  \log \mathbb{E}\, \mathcal{N}_\mathrm{fp}$, is tight for the core model and for $L=2$. For instance, for the two layer case we derive in the Supplementary Material (SM-A):
\begin{align}
& S_\mathrm{fp} =  \lim_{N \to \infty} \frac{1}{2N}  \log \mathbb{E}\, {\cal N}_\mathrm{fp}= \nonumber \\
& = \frac{\log 2}{2}  +\frac{1}{2}\log \left[ H\left(-J_D-\lambda \right)^2 +H\left(-J_D+\lambda \right)^2\right]
 \end{align}
where \( H(x) = \frac{1}{2} \operatorname{erfc}\left( \frac{x}{\sqrt{2}} \right) \). The entropy $S_\mathrm{fp}$ is an increasing function of $J_D $ and $\lambda$, it vanishes when \( J_D = \lambda= 0 \), and goes to $\log 2$ in the limit \( J_D, \lambda \to \infty \), where all configurations become fixed points. The core model is recovered by setting $\lambda=0$. A replica computation of the same quantity confirms that typical fixed points are isolated. According to the overlap gap property (OGP) \cite{gamarnik2021overlap}, this implies that stable algorithms (which include the network dynamics as a specific case)  cannot reach such fixed-point configurations in sub-exponential time.

 While these analytical calculations correctly give the behavior of dominant configurations, they neglect the role of subdominant configurations (with an entropy smaller than $S_\mathrm{fp}$)
which are not isolated and can be  attractive for some dynamical processes and algorithms ~\cite{baldassi_prl_2015}. In fact, extensive simulations with several algorithms \cite{furthcoming2025} show that there exist algorithm-dependent threshold values of $J_D$ beyond which fixed points can be reached very rapidly. For the core model, Belief Propagation with reinforcement \cite{baldassi_pnas_2016} appears to be able to find fixed points with $J_D$ as low as $0.35$ for large $N$ in a small (sublinear) number of iterations; likelihood optimization \`a la ref.~\cite{baldassi2018role_of_stochasticity}  requires $J_D \gtrapprox 0.5$; simple iteration of the update dynamics, on which we will focus in this paper, converges for $J_D \gtrapprox 0.8$.
(see SM-A for the scaling bound $J_D \lessapprox\sqrt{2 \ln N}$).
The dynamics converge to the RM, an extended connected cluster with a high density of solutions and a branching structure \cite{baldassi2021unveiling,annesi2023star}. It can be studied by a  more refined analytical approach where one biases the Gibbs measure toward fixed points surrounded by a large number of neighboring solutions \cite{baldassi_prl_2015}; this ``local entropy approach'' is explained in the ``End Matter'' section and detailed in SM-B.



The existence of the RM is a prerequisite for the convergence of local dynamical algorithms and underlies the effectiveness of our learning scheme. Its geometry combines flexibility and stability: the RM forms a dense, connected cluster of fixed points that are robust to substantial perturbations. As discussed in the Supplementary Material, perturbing a fixed point by flipping a large fraction of its units typically drives the dynamics toward another fixed point in its immediate neighborhood, at a Hamming distance of only a few percent. Learning further enhances this robustness by stabilizing the reached configuration, which remains stable even after the transient output feedback is removed. This persistence of task-relevant configurations provides a natural mechanism for generalization.

Moreover, the persistence of the state after the removal of the output signal, made possible by the geometrical properties of the RM, offers a simple and biologically viable protocol to implement a supervised learning scheme via local Hebbian-like weight updates, as described in the next section. This is very different from standard learning schemes which implausibly require neurons to compare their actual state with a hypothetical state reached in absence of the supervisory signal. Alternative proposals to circumvent this issue are generally more complicated, e.g. relying on contrastive temporal comparisons \citep{saglietti2018statistical} or more elaborate learning rules \citep{alemi2015three}.

\textbf{Learning.} For simplicity, we shall describe our supervised learning protocol on the Core Module; the generalization to multilayer architectures is straightforward and can be found in the SM-C.

Assume that we are provided with a set of \(P\) patterns with their associated labels \( \{ x^\mu , y^\mu \}_{\mu=1}^P\) with \(x^\mu \in \mathbb{R}^D\) and \( y^\mu \in \{-1, 1\}^C\).
We consider two projection matrices \(W^\mathrm{in} \in \mathbb{R}^{N \times D}\), \(W^\mathrm{back} \in \mathbb{R}^{N \times C}\) that mediate the influence of the input and label onto the network state, as well as a readout matrix \(W^\mathrm{out} \in \mathbb{R}^{C \times N}\) that outputs a prediction given the state of the network (see Fig. \ref{fig:models}). These matrices can be initialized and updated similarly to \(J\).

During training, pairs \(x^\mu, y^\mu\) are presented one at a time. The network evolves in parallel by aligning neurons to their local field:
\begin{align}
    s_i \leftarrow \sign \Bigg( \sum_{j=1}^N J_{ij} \, s_j 
    &+ \lambda_x \sum_{k=1}^D  W^\mathrm{in}_{ik} x^\mu_k \notag \\
    &+ \lambda_y \sum_{c=1}^C W^\mathrm{back}_{ic} y^\mu_c \Bigg) \;,
    \label{eq:evolution}
\end{align}
where \(\lambda_y\) controls whether the output influences the dynamics.

We initialize the network in a neutral state \(s=0\). Initially, only the input affects the dynamics for a few \emph{warmup} steps (set \(\lambda_y = 0\) in \eqref{eq:evolution}; with \(L\) layers we use \(L\) warmup steps). Then, the supervision kicks in (\(\lambda_y > 0\)) and the network reaches a fixed point \(s'\). From here, the transient supervision is removed (\(\lambda_y = 0\)) and the network reaches a second fixed point \(s^*\). 
To stabilize \(s^*\), we apply a simple local plasticity rule that increases the stability margin \(s_i^* \cdot h_i\) of each neuron \(i\) by \(\eta > 0\) if it falls below a positive threshold \(\kappa\):
\begin{align}
    \label{eq:hidden-field}
    h_i &= \sum_{j=1}^N J_{ij} \, s_j^* + \lambda_x \sum_{k=1}^D  W^\mathrm{in}_k x^\mu_k \\
    J_{ij} &\leftarrow J_{ij} +  \eta \, s_i^{*} \, s_j^{*} \; \mathds{1} (s_i^{*} \cdot h_i \le \kappa)\mathds{1}(J_{ij}\neq 0)
    \label{eq:plasticity_rule}
\end{align}
The projections \(W^\mathrm{in}\), \(W^\mathrm{out}\) and \(W^\mathrm{back}\) can be learned using exactly the same principle: \(W^\mathrm{in}\) and \(W^\mathrm{back}\) learn to increase the margin \(s_i^* \cdot h_i\) of each hidden neuron, while \(W^\mathrm{out}\) learns to increase the margin of the prediction \(y_c^\mu \cdot \hat{y}_c\), where \(\hat{y} = W^{out} s^*\).
Note that the training of the readout matrix \(W^\mathrm{out}\) does not affect the learning dynamics of the rest of the network, and could thus be carried out with any other regression algorithm learning to map \(s^*\) into \(y^\mu\).

During inference, the network evolves under the sole influence of the input, and upon convergence to a state $\hat{s}$ we
compute the prediction $\hat{y}= W^{out} \hat{s}$.

In practice, already after a few learning steps, convergence of the dynamics typically requires no more than 5-10 update steps. Furthermore, using the synchronous dynamics \eqref{eq:evolution} allows an efficient vectorized implementation of our method on modern hardware, especially considering batches of data in parallel (see SM-C).

\textbf{Performance Benchmarks.} 
\label{sec:experiments-manifold}
We benchmark the performance of our algorithm (see SM-D) against a two-hidden-layer multilayer perceptron (MLP) trained with backpropagation, on a set of image classification benchmarks: MNIST \cite{lecun1998gradient}, Fashion-MNIST \cite{xiao2017fashion}, Entangled MNIST (a more challenging MNIST variant, in which MNIST features are randomly projected onto a reduced 100-dimensional space and binarized, see SM-D), and the Tiny ImageNet subset of ImageNet \cite{deng2009imagenet} preprocessed by a pretrained convolutional backbone. The code to reproduce our experiments is released at ref.~\cite{github-darnax}, 
and raw experimental data are available from the authors upon reasonable request.

We consider the core module, fixing the sparsity coefficients to $\rho_J=0.01$, $\rho_{W_{in}}=0.1$ and $\rho_{W_{out}} = 1.0$ and varying the number of neurons $N$ linearly. To ensure a fair comparison, we also vary the width of the MLP to exactly match the total number of learnable parameters. Furthermore, since our model operates with binary neurons, also the MLP employs tanh nonlinearities followed by a sign function, to binarize activations; during the backward pass, gradients are computed ignoring the sign, via straight-through estimation \cite{bengio2013estimating}.

\begin{figure}[h!]
    \centering
    \includegraphics[width=1.0\linewidth]{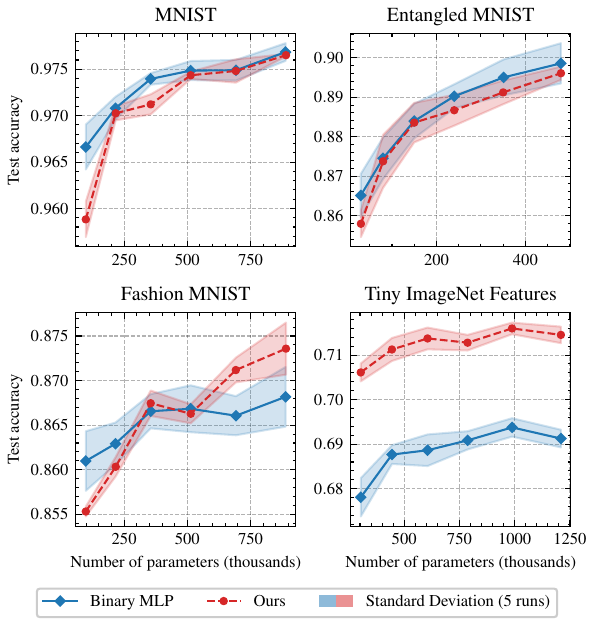}
    \caption{Test accuracy on benchmark datasets as a function of the number of trainable parameters, controlled by the layers width.
    Our model (red) is compared to a binarized 3-layer perceptron trained with backpropagation (blue). For each value of trainable parameters, the average test accuracy over 5 distinct runs is plotted, together with its standard deviation.}
    \captionsetup{format=plain, font=small, labelfont=bf}
    \label{fig:performance-benchmarks}
\end{figure}

Fig.~\ref{fig:performance-benchmarks} shows  that our approach achieves performance comparable to that of backpropagation-based training in terms of test accuracy.

We also investigate the impact of depth in the multilayer chain model, focusing on the challenging task of hetero-associative learning (see SM-D for task description and figures). We observe a linear dependence of the hetero-association capacity on network width, with a coefficient that increases monotonically with depth. Furthermore, the overlap of the internal representations with the target output increases gradually with depth. We expect depth to play an even more significant role when combined with suitable inductive biases in the connectivity, and leave a systematic exploration of this to future work.

Finally, we assess more closely the ability of the proposed algorithm to perform feature learning. In SM-D we compare the performance of the core module with two simple baselines that train a linear readout on frozen features (random projection and reservoir computing \cite{Jaeger2001echostate}), matching the dimensionality of the representations. Our algorithm learns features that allow to consistently outperform these baselines in terms of training and test accuracy on Entangled MNIST, with especially pronounced gains when using low-dimensional features.



\label{sec:experiments-manifold}

\begin{figure}[h!]
    \centering
    \includegraphics[width=1.0\linewidth]{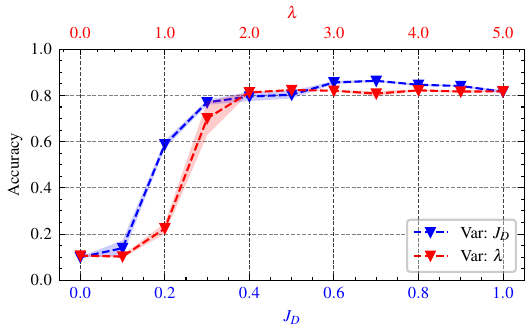}
    \caption{Test accuracy of a 2-layer chain model on Entangled MNIST as function of $\lambda$, $J_D$. For each curve, one parameter is varied while the other is kept equal to 0.}
    \captionsetup{format=plain, font=small, labelfont=bf}
    \label{fig:jd-lambda-duality}
\end{figure}

\textbf{On the role of the RM.}
The importance of the RM accessibility and its geometry for learning is confirmed by several sets of experiments.
In the first experiment (SM-E), we measured how the performance of our algorithm depends on the couplings $J_D$ and $\lambda$, using a 2-layer chain model where, additionally, both layers interact  with the input and the output independently. This can be thought of as 2 coupled and independently initialized core modules whose predictions are ensembled. We vary separately each parameter, fixing the other to 0, and we measure classification performance on the Entangled MNIST test set after a fixed number of epochs.  Fig. \ref{fig:jd-lambda-duality} shows that the network is unable to learn the task when \(J_D\) and \(\lambda\) are too small, but its performance suddenly and drastically improves as \(J_D\) or \(\lambda\) increase. Crucially, because both copies of the core module independently interact with input and output, the information propagates from the input through both layers and to the output even with \(J_D = \lambda = 0\). This finding thus corroborates that the sudden performance boost is rooted in the dynamical properties of the model, and that it is directly linked with the onset of the RM. 

We also study the impact of the symmetry in the coupling matrix \(J\) on performance, by considering variants of the core module that initialize \(J\) symmetrically and, optionally, use variants of the plasticity rule that preserve the symmetry (see SM-F). 
With \(J_D = 0\), the dynamics of the symmetric model descends a Lyapunov function and is convergent. However, it turns out that introducing and even preserving symmetry in \(J\) has no significant effect on model performance, compared to the asymmetric case. Furthermore, the symmetric variants also exhibit a drastic performance boost as \(J_D\) increases, despite the dynamics already being convergent with \(J_D = 0\). This shows that the critical aspect of the RM for learning is not only its accessibility but also its geometric structure, the details of which are explored in SM-G.

\textbf{Conclusion.}
We have proposed a new scheme for asymmetric recurrent neural networks based on the dynamical stabilization of accessible representation manifolds. This distributed gradient-free mechanism, backed by statistical physics theory, is local both in space and in time, and it provides a viable principle for learning in recurrent neural networks, which can even compete with the state-of-the-art gradient-based algorithms. 

\textbf{Acknowledgments}. We thank Nicolas Brunel for very stimulating discussions. This work was supported by the
PNRR-PE-AI FAIR project funded by the NextGeneration EU program.

\section{End Matter\label{sec:endmatter}}

We explain here the  ``Local Entropy approach'', an analytic method   showing the existence of the RM in the system before learning takes place. It uses an effective energy density $ \mathcal{E}(\tilde{s},d)$ relative to a reference configuration of the neurons \(\tilde{s}\),  and the associated partition function defined as
\begin{equation}
\mathcal{E}(\tilde{s},d) = -\frac{1}{N} \log \mathcal{N}(\tilde{s}, d)\ ;\ Z(d,y)=\sum_{\tilde s} e^{-y N \mathcal{E}(\tilde{s},d)}
\end{equation}
where \(\mathcal{N}(\tilde{s}, d)\) is the number of fixed points \(s\) at normalized Hamming distance \(d\) from \(\tilde{s}\), and the parameter \(y\) plays the role of an inverse temperature, controlling the reweighting of the local entropy contribution. 
The corresponding free energy density 
\begin{equation}
\Phi(d,y)=-\lim_{N\to\infty} \frac{1}{N y} \; \log  Z(d,y)
\label{eq:free_entropy}
\end{equation}
allows to reveal sub-dominant regions with highest  density of fixed points in the limit of large \(y\) and small \(d\).
Relevant quantities are the average local entropy, i.e. the  internal entropy density of the RM, $S_I(d, y)\equiv \langle {\cal E} \rangle = - \frac{\partial}{\partial y} \left[y \Phi(d,y)\right]$, and the log of number of such regions, called the ``external entropy density'', $S_E(d, y) = - y \,  (\Phi(d, y) + S_I(d, y))$. At fixed $d$, $S_I$ is an increasing function of $y$ and $S_E$ is a decreasing function. The case  \( y = 1 \) reduces to the computation of the entropy surrounding typical fixed points, which in our case would be the isolated ones. Considering values of \( y > 1 \) implies that we are exploring subdominant regions or out-of-equilibrium states. 

As explained in ref.~\cite{baldassi_prl_2015}, the  signature of the existence of an accessible  RM
is that, at the largest value of $y$ for which the external entropy $S_E(d,y) \ge 0$, call this $y^*(d)$,  the internal entropy  $S_I(d, y^*(d))$ is a monotonically growing function of $d$.

We evaluate Eq.~\ref{eq:free_entropy} in the large $N$ limit, using the replica method under the replica symmetric (RS) Ansatz. As discussed in ref.~\cite{baldassi2020shaping}, this computation can be formally interpreted as the  one-step replica symmetry breaking (1-RSB) calculation of the original problem of computing the entropy of fixed points, where now the Parisi parameter \( m \) must be  identified with \( y \), and the internal overlap \( q_1 \) is used as a control parameter to fix the region's diameter via \( d = (1 - q_1)/2 \).

The asymptotic  expression  for the free energy and related entropies can be derived for an arbitrary number of layers, see SM-B. However, solving the saddle point equations poses difficult numerical challenges, and we limit our analytical study to the case of two layers,  which is enough to display the role of  both  $J_D$ and $\lambda$. The expression for the free energy density reads:
\begin{align}
-y&\Phi(d,y)  = 
-\hat q_1
+ (1 - y) \, \hat q_1 q_1 
+ \frac{1}{y} \! \log \! \Bigg\{
    \!\! \int \!\! Dz\, Dz' Dt\, Dt' \nonumber \\
& \quad \Bigg[
    \sum_{s, s'\in \{\pm 1\}}
    e^{\sqrt{\hat q_1} (s t + s' t')}
    H\left(
        \frac{-J_D + \lambda\, s s' - s \sqrt{q_1} z}{\sqrt{1 - q_1}}
    \right)
    \nonumber \\
& \quad\quad 
    \times H\left(
        \frac{-J_D + \lambda\, s s' - s' \sqrt{q_1} z'}{\sqrt{1 - q_1}}
    \right)
\Bigg]^y
\Bigg\}
\end{align}
where $Dz,  Dz', Dt, Dt'$ are Gaussian measures with mean 0 and variance 1, and   $\hat q_1$ is determined by  the saddle point equation $\partial \Phi/\partial\hat q_1=0$ (see SM-B). The results for the core and multilayer chain models are as follows.

Consider the single core module. The free energy density is recovered by setting $\lambda=0$ and dividing by 2 (to account for the number of neurons). Solving the saddle-point equations reveals that, for all \( J_D > 0 \) and \( d \), there exists a value of \( y \) beyond which the external entropy \( S_E(d, y) \) becomes negative. This is unphysical and indicates a breakdown of the RS assumption, signaling a phase transition. To identify the critical values of \( J_D \), we first compute, for each value of \( d \), the value \( y^* \) at which the external entropy vanishes, and next seek the value of $J_D$ below which the internal entropy becomes a non-monotonic function of the distance. We find that  the phase space separates in two regions.   
For all \( J_D < J_{Dc}\simeq 0.07 \), the internal entropy is non-monotonic. The value of \( d = (1 - q_1)/2 \) that maximizes the entropy coincides with the self-overlap of subdominant  states, given that the associated value of \(y^* \) exceeds one.
The actual values of \( d \) are extremely small, suggesting that finding  fixed points belonging to such states would still be hard for local algorithms. 
For all \( J_D > J_{Dc} \), the internal entropy increases monotonically with \( d \), eventually saturating at a plateau equal to the entropy of typical solutions. This behavior indicates the existence of an extensive cluster in configuration space where fixed points coalesce giving rise to the RM geometry.

\begin{figure}[t]
    \centering
    \includegraphics[width=1 \linewidth]{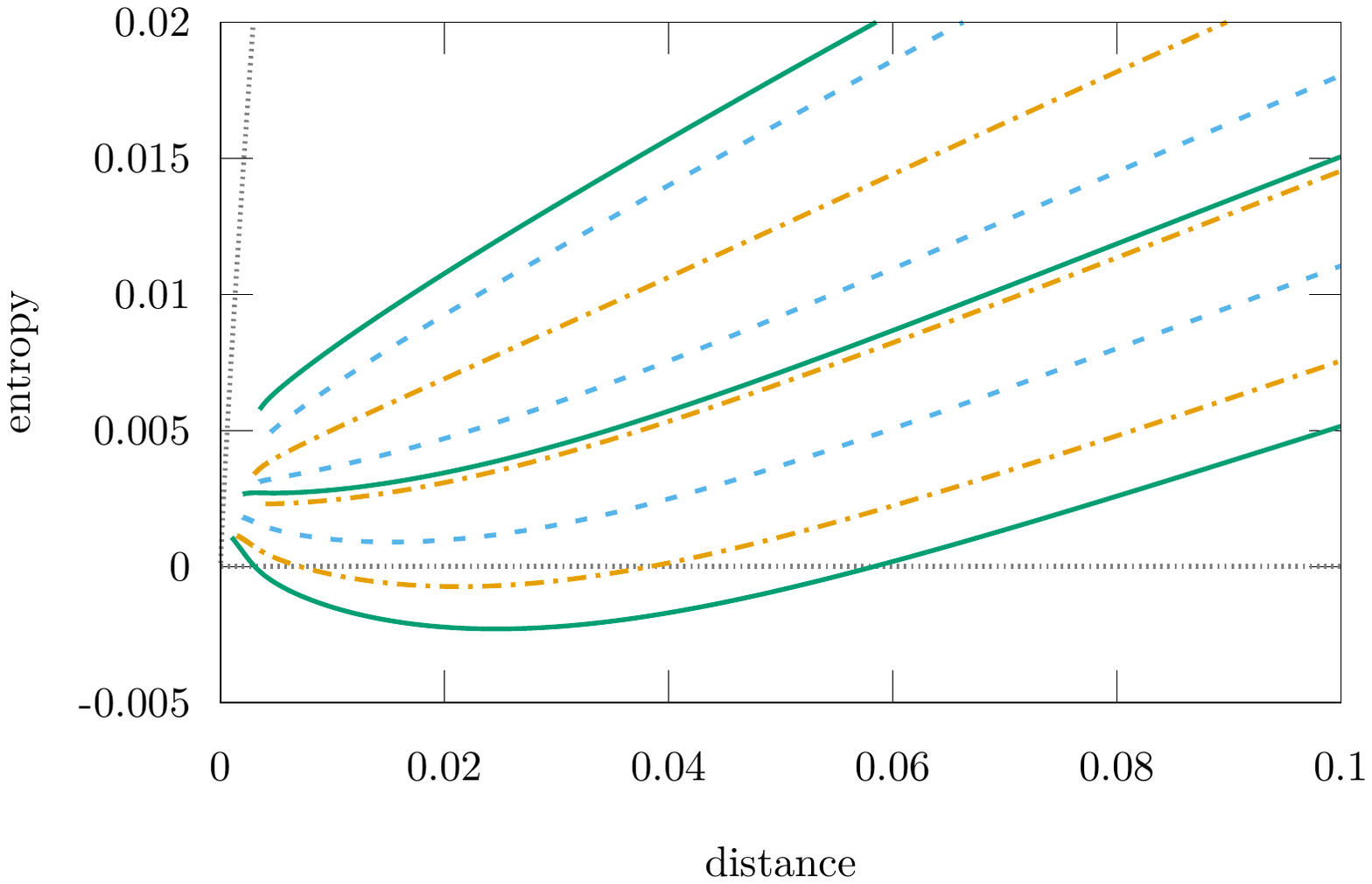}

    \caption{Internal entropy (divided by $L$) for the Core Model (solid green) and two-layer model (dashed/dot-dashed blue/orange). Core Model: $J_D=0.05,0.07,0.1$ (bottom to top). Two-layer: blue ($J_D=0.05$) with $\lambda=0.2,0.3,0.4$ (bottom to top); orange ($J_D=0$) with $\lambda=0.4,0.45,0.5$ (bottom to top). All curves vanish at distance $0$ (not shown due to numerical limits). In all cases, increasing $J_D$ or $\lambda$ drives the OGP–RM transition.}
    \captionsetup{format=plain, font=Large, labelfont=bf}
    \label{fig:local_entropies}
\end{figure}

The same analysis can be extended to the multilayer model. Focusing on the case of $L=2$, we find that the role of \(\lambda\) closely parallels that of \(J_D\): a positive sparse coupling $\lambda$ between different subsystems induces the emergence of a RM.
Specifically, for small values of \(J_D\), including zero, a critical value of \(\lambda \lessapprox 0.45\) exists at which the extended high-density cluster of fixed points emerges.   Fig. \ref{fig:local_entropies} shows the behavior of the local entropy for the two cases.

\bibliography{bibliography}
\onecolumngrid
\newpage
\section{Supplementary Material}
\label{appendix}

\subsection{A- The entropy of fixed points}
\input{annealed}

\subsection{B- Local Entropy}
\input{local_entropy}

\subsection{C- Learning algorithm and its extensions}
\input{training-variants}
\subsection{D- Experimental details}
\input{benchmarks-hyperparameters-and-details}
\subsection{E- Geometry of the RM}
\input{importance-stable-manifold}
\subsection{F- Evolution of symmetry}
\input{evolution-symmetricity}
\subsection{G- Landscape analysis}
\input{landscape_stability}

\subsection{H- Evolution of internal states}
\input{evolution-of-internal-states}

\end{document}

%% file: annealed.tex
We consider a neural network made of a chain of $L$  core modules.
Each core module  with $\ell\in\{2,...,L-1\}$ is coupled to the module $\ell-1$ and the module $\ell+1$ through a positive couplings $\lambda_L\geq 0$ and $\lambda_R \geq 0$. For the sake of simplicity, in the main text we discussed just the case $\lambda_L=\lambda_R=\lambda$. 
We are interested in the number of fixed points of the  dynamics.
Let us denote the local fields at layer $\ell$ by:
\begin{align}
    h_i^\ell&= \sum_j J_{ij}^\ell s_{j}^\ell \\
\end{align}
For $L \geq 2$, the fixed points conditions are expressed as :
\begin{align}
     \forall \ell \in\{2,...,L-1\}:&
    \ \ s_{i}^\ell h_{i}^\ell +J_D+\lambda_L s_{i}^{\ell-1}s_{i}^{\ell}+\lambda_R s_{i}^{\ell}s_{i}^{\ell+1}>0
    \nonumber\\
    \text{For} \ \ell=1:
    &\ \ s_{i}^\ell h_{i}^\ell +J_D+\lambda_R s_{i}^{\ell}s_{i}^{\ell+1}>0
     \nonumber\\
    \text{For} \ \ell=L:
    &\ \ s_{i}^\ell h_{i}^\ell +J_D+\lambda_L s_{i}^{\ell}s_{i}^{\ell-1}>0
    \label{Eq:fpeqs}
\end{align}
If $L=1$ there are no $\lambda$-couplings and the fixed point condition is just
\begin{equation}
     s_{i}^1 h_{i}^1 +J_D >0
    \label{Eq:fpeqs_1}
\end{equation}
The number of fixed points is given by
\begin{align}
    {\cal N}_\mathrm{fp}= \sum_{s} \prod_{\ell=1}^{L-1} \prod_{i=1}^N \theta\left(u_{i}^\ell(s)\right)
\end{align}
where the $\sum_s$ denotes a sum over all the internal representations $\{s_{i}^\ell\}$ for $\ell\in\{1,...,L\}$, while $u_i^\ell(s)$ are the conditions appearing in Eqs. (\ref{Eq:fpeqs}) and (\ref{Eq:fpeqs_1}).
The entropy of fixed points is defined as 
\begin{align}
 S_\mathrm{fp} =  \lim_{N \to \infty} \frac{1}{LN}  \log  {\cal N}_\mathrm{fp}
 \end{align}
 
\textbf{Annealed average}

We start the study by considering the  annealed average of this number of fixed points over the choice of the couplings $J^\ell$  with their respective Gaussian measures. It proceeds as follows.
We first notice that, when averaging over the $J$, the variables $h_{i}^\ell$ are iid Gaussian of mean zero and variance unity.
Defining 
\begin{align}
    H(x)=\int_x^\infty \frac{dt}{\sqrt{2\pi}}e^{-t^2/2}\ ,
\end{align}
the annealed average of the number of fixed points is $ \overline {\cal N}_{fp}= z^N$ where the partition function $z$ is 
\begin{align}
    z=
    \sum_{s^1,...,s^{L}}H\left(-J_D-\lambda_R s^{1}s^{2} \right) \prod_{\ell=2}^{L-1}
    H\left(-J_D-\lambda_L s^{\ell-1}s^\ell-\lambda_R s^{\ell}s^{\ell +1}\right) 
    H\left(-J_D-\lambda_L s^{L-1}s^{L}\right)
\end{align}    
This partition function can be computed  by transfer matrix. Note that the weight is invariant by flipping all the spins. So we can compute $z_+$ which is the partition function when we fix $s_1=1$ and we have $z=2 z_+$. 

{\bf Transfer matrix computation} 

We define the vector
 \begin{align}
     f_k(s^k,s^{k+1})= \sum_{s^1,...,s^{k-1}}
     H\left(-J_D-\lambda_R s^{1}s^{2} \right)\prod_{\ell=2}^{k}
     H\left(-J_D-\lambda_L s^{\ell-1}s^\ell-\lambda_R s^{\ell}s^{\ell +1} \right) 
 \end{align}
 which satisfies the recursion relation
 \begin{align}
     f_{k+1}(s^{k+1},s^{k+2})= \sum_{s^{k}}f_k(s^k,s^{k+1}) 
     H\left(-J_D-\lambda_L s^{k}s^{k+1}-\lambda_R s^{k+1}s^{k+2} \right) 
     \label{eq:rec}
 \end{align}
 with the initial condition 
\begin{align}
     f _1(s_1,s_2)=  H\left(-J_D-\lambda_R s^{1}s^{2} \right)
\end{align}
 We can express the annealed entropy associated with the average number of fixed points as:
\begin{align}
S_a=
   \frac{1}{N L}\log \overline {{\cal N}_\mathrm{fp}}= \frac{1}{L}\log 2+ \frac{1}{L}\log 
    \sum_{s^{L-1},s^{L}} f_{L-1}(s^{L-1},s^{L}) \; H\left(-J_D-\lambda_L s^{L-1}s^{L} \right)
   \label{N_annealed}
\end{align}
We introduce a four dimensional vector $\vec f_k$ defined as
 \begin{align}
     \vec f_k=\left(
\begin{array}{c}
f_k(-1,-1) \\
f_k(-1,1) \\
f_k(1,-1) \\
f_k(1,1) \\
\end{array}
\right)\ .
 \end{align}
 Then the recursion (\ref{eq:rec}) can be expressed as $\vec f_{k+1}=M_{k+1}\vec f_k $ where $M_{k+1}$ is the $4\times 4$ matrix:
 \begin{align}
   \left(
\begin{array}{cccc}
H_{--}& 0 & H_{+-}& 0\\
H_{-+}& 0 & H_{++}& 0\\
0&H_{++}& 0 & H_{-+}\\
0& H_{+-}& 0 & H_{--}
\end{array}  
\right)
 \end{align}
 where 
 \begin{align}
     H_{\tau_1 \tau_2}= H\left(
     -J_D+\tau_1\lambda_L +\tau_2 \lambda_R\right)\ ,
 \end{align}
 with $\tau_1,\tau_2\in\{\pm 1\}$.
The initial vector is 
\begin{align}
     \vec f_1=
     \left(
\begin{array}{c}
0 \\
0 \\
H\left(-J_D+\lambda_R \right)\\
H\left(-J_D-\lambda_R \right) \\
\end{array}
\right)
\end{align}
 The average number of fixed points is given by (\ref{N_annealed}) where $\vec f_{L-1}=M^{L-2}\vec f_1$.
 Shallow networks are special cases :
\begin{align}
     L=1:\ \  \frac{1}{N}\log \overline {\cal N}_\mathrm{fp}=& \log 2 + \log   H\left(-J_D \right)\\
     L=2:\ \  \frac{1}{2N}\log \overline {\cal N}_\mathrm{fp}=& \frac{1}{2} \log 2  + \frac{1}{2}\log \left[   H\left(-J_D-\lambda_R \right)H\left(-J_D -\lambda_L\right) + \right. \nonumber \\
     &      \left. H\left(-J_D+\lambda_R \right)
     H\left(-J_D +\lambda_L\right)
     \right]
 \end{align}

\begin{figure}[h!]
    \includegraphics[width=.4\textwidth]{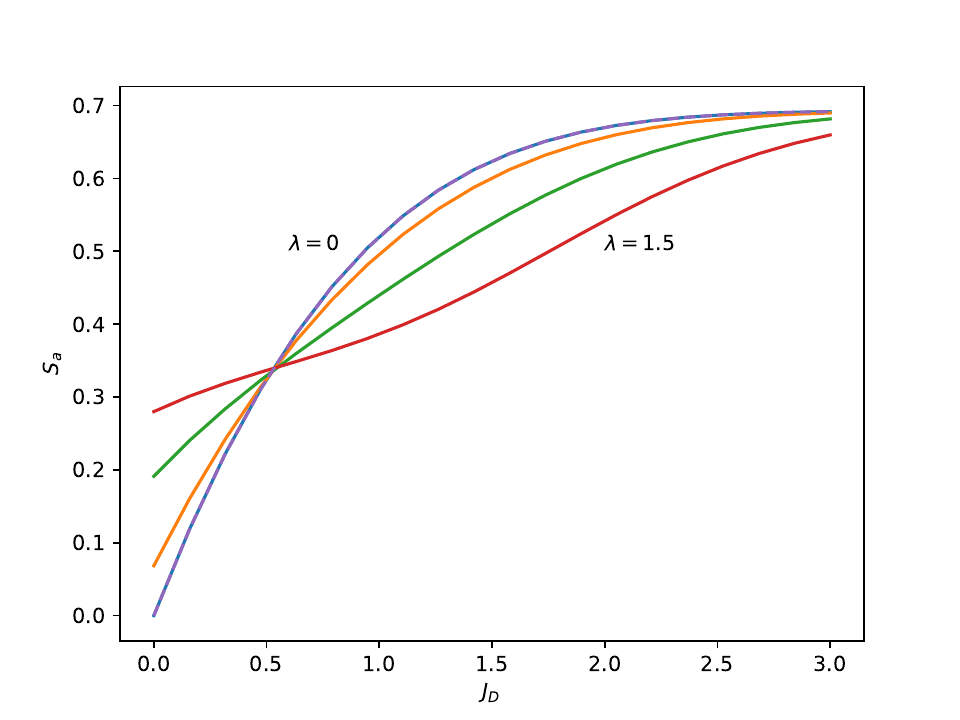}
    \includegraphics[width=.4\textwidth]{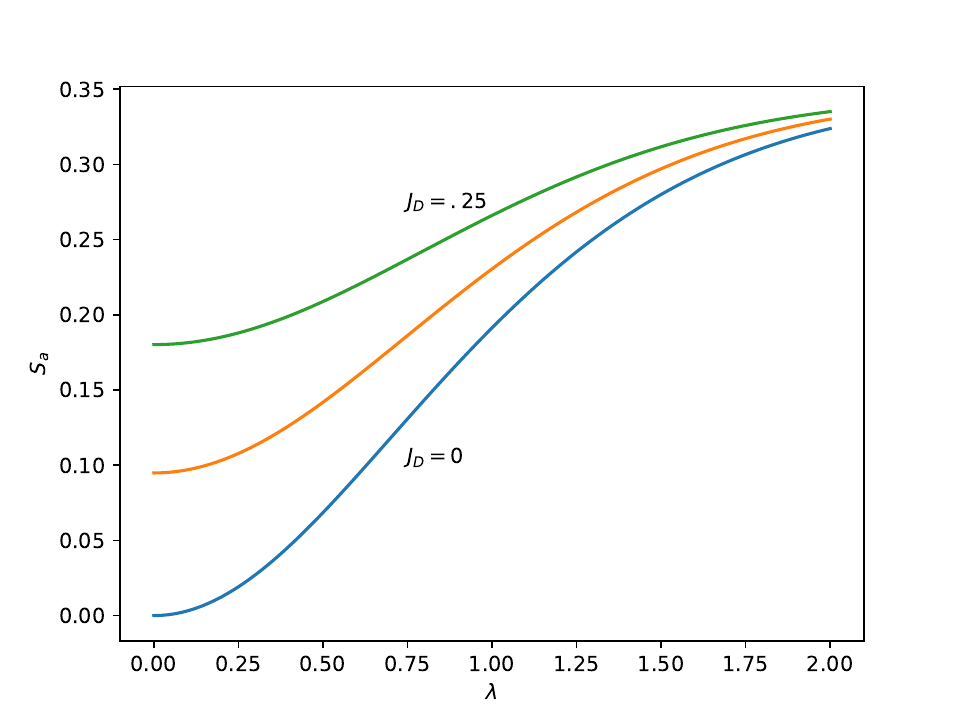}
  \caption{Annealed entropy $S_a$ of the number of foxed points, in the case of two core modules, $L=2$, and $\lambda_L=\lambda_R=\lambda$. Left: $S_a$ is plotted versus $J_D$, for $\lambda=0,0.5,1,1.5$. Right: $S_a$ is plotted versus $\lambda$, for $J_D=0,0.125,0.25$}
  \label{fig:annealed_entropy}
\end{figure}

The result for $L=2$ is shown in Fig. \ref{fig:annealed_entropy}. Within this annealed approximation, one sees 
that the use of nonzero values of $\lambda_L,\lambda_R$ plays a role similar to $J_D$ in stabilizing fixed points.  
\\

\textbf{Tightness of the annealed average} 

The annealed average that we have just performed approximates the quenched average  $\overline \log {\cal N}_\mathrm{fp}$ by its first moment $ \log \overline {\cal N}_\mathrm{fp}$.
Let us show that, in  the $L=1$ case, this first moment bound is tight. Let's give here a brief account of calculations for the core model ($L=1$). A detailed discussion in more general cases will be presented in [21].

A configuration $\mathbf s$ is a fixed point if $s_i(\sum_{j\neq i}J_{ij}s_j + J_D s_i)= h_i s_i+J_D >0$ for all $i$. As discussed above, for fixed $\mathbf s$, we have that the local fields are Gaussian
\[
h_i:=\sum_{j\neq i} J_{ij}s_j \sim \mathcal N\!\Big(0,\frac{N-1}{N}\Big),\quad
\]
and the probabilities of satisfying the fixed point conditions are
\[
{\cal P}\big(s_i h_i+J_D>0\big)=H\!\Big(- \tfrac{J_D}{\sigma_N}\Big),
\]
with $\sigma_N^2=(N-1)/N\to1$.  Since rows are independent, we have 
\begin{equation}
\label{eq:EZ}
\mathbb{E} [Z]=\sum_{\mathbf s}\prod_{i=1}^N H\!\Big(-\tfrac{J_D}{\sigma_N}\Big)
= \Big(2\, H\!\big(-\tfrac{J_D}{\sigma_N}\big)\Big)^{\!N}
\;\xrightarrow[N\to\infty]{}\;
\big(2\,H(-J_D)\big)^{N}.
\end{equation}
Upon introducing the overlap  $q=\frac{1}{N}\sum_i s_i s'_i\in[-1,1]$  between pairs of configurations $s$ and $s'$, and summing over $q$, the  averaged second moment of $Z$ can be written as
\[
\mathbb{E} [Z^2]= \sum_{k=0}^N 2^N \binom{N}{k}\,
F(J_D, q_k)^{k}\,
F(J_D,-q_k)^{N-k},
\label{eq:EZ2}
\]
where $ q_k=\tfrac{2k-N}{N}$ and
\[
F\big(J_D, x \big)=\int_{-J_D}^\infty \frac{e^{-t^2/2} }{\sqrt{2 \pi}} dt \;  H\left(-\frac{x t+J_D}{\sqrt{1-x^2}}\right)
\]
To write $\mathbb{E}[Z^2]$ in the limit $N \to \infty$, we need to solve a saddle point equation. Setting $x=k/N$ and $q=2x-1$, $\mathbb{E}[Z^2]$ has a  large-$N$ exponent
\[
\frac{1}{N}\log \mathbb{E}[Z^2] \;=\; \sup_{q\in[-1,1]}
\Big\{\log 2 + H_b\big(\tfrac{1+q}{2}\big)
+ \tfrac{1+q}{2}\log F(J_D, q)
+ \tfrac{1-q}{2}\log F(J_D, -q)\Big\},
\]
where $H_b(p)=-p\log p-(1-p)\log(1-p)$ is the binary entropy. This exponent is even in $q$ and  the maximum is attained at $q=0$. Therefore, to the leading exponential order,  one finds:
\[
\frac{1}{N}\log \mathbb{E}[Z^2]\ =\ 2\log\!\big(2\,H(-J_D)\big)\ =\ \frac{2}{N}\log \mathbb{E}[Z]\ .
\]
It follows:
\[
\frac{\mathbb{E}[Z^2]}{\mathbb{E}[Z]^2}\ \;\xrightarrow[N\to\infty]{}\; 1
\]
with a vanishing relative variance $\frac{\mathrm{Var}(Z)}{\mathbb{E}[Z]^2}\ \longrightarrow\ 0 $.
Hence the annealed estimate $\,\mathbb{E}[Z]\simeq \big(2 H(-J_D)\big)^N\,$ is \emph{self-averaging} and gives a tight  exponential order bound in the case $L=1$. \\

\textbf{Convergence bound for the update dynamics} 

One can easily obtain a convergence threshold bound for the simple update dynamics by asking for a value of $J_D$ large enough that none of  $N$ randomly fluctuating local fields  is likely to exceed the flipping threshold.

Let $h_1,\dots,h_N \sim \mathcal N(0,1)$ be the i.i.d. local fields. Using a union bound and the Gaussian tail bound,
\begin{equation}
\mathbb P\!\left(\max_{1\le i\le N} |h_i| > J_D \right)
\le \sum_{i=1}^N \mathbb P(|h_i| > J_D)
\le 2N e^{-J_D^2/2}.
\end{equation}
Requiring this probability to be $O(1)$ yields
\begin{equation}
2N e^{-J_D^2/2} \sim 1
\;\;\Rightarrow\;\;
J_D^2 \sim 2\ln N,
\end{equation}
hence
\begin{equation}
J_D = \sqrt{2\ln N}.
\end{equation}\\

\textbf{Replica symmetric computation}\\

A complementary approach consists in computing the entropy of fixed points using the replica method. \\

\textbf{General expression}
 
The $n$-th power of the number of fixed points is given by
\begin{equation} 
\overline{{\cal N}_\mathrm{fp}^n}= \sum_{s_a} \prod_{\ell=1}^{L} \prod_{i} \prod_{a=1}^n \theta\left(u_{i} ^{\ell }(s_a)\right)
\end{equation}
where the $\sum_{s_a}$ denotes a sum over all the internal
representations $\{s_{ia}^{\ell}\}$ for $\ell\in\{1,...,L\}$. When sampling the $J$, the local fields $
  h_{ia}^{\ell }=  \sum_j J_{ij}^\ell s_{ja}^{\ell} $
have a joint Gaussian distribution with mean zero. They are independent for each $i,\ell$ and their covariance in replica space is 
\begin{align}
\mathbb{E} \left[ h_{ia}^{\ell } h_{jb}^{m  }\right]= \delta_{i j}\delta^{\ell m} q^\ell_{ab}
\end{align}
where
\begin{align}
    q^\ell_{ab}=\frac{1}{N}\sum_j s_{ja}^{\ell }s_{jb}^{\ell }\ .
\end{align}

Introducing this replica overlap $ q^\ell_{ab}$ in each layer $\ell$ and the corresponding Lagrange multiplier 
$\hat  q^\ell_{ab}$, we find the expression:
\begin{align}
    \overline{{\cal N}_\mathrm{fp}^n}&= \sum_{s} \int \prod_{a<b} dq_{ab} d\hat q_{ab}
    e^{-N\sum_\ell \sum_{a<b} \hat q^\ell_{ab}q^\ell_{ab}+\sum_{\ell,i,a,b}\hat q^\ell_{ab}s_{ia}^\ell s_{ib}^\ell }\nonumber\\
  &\prod_{i\ell} \left[
    \mathbb{E}_{\{h_a\}\sim\mathcal{N}(0,q^\ell _{ab})} \;\prod_{a=1}^n \theta\left(s_{ia}^\ell h_{a}+J_D+\lambda_L s_{ia}^{\ell-1}s_{ia}^{\ell}+\lambda_R s_{ia}^{\ell}s_{ia}^{\ell+1}\right)
  \right] 
\end{align}
where the product over $\ell$ runs from $\ell=1$ to $\ell=L$, and it is understood that the term $ \lambda_L
s_{ia}^{\ell-1}s_{ia}^{\ell}$ is absent when $\ell=1$, and the term $
\lambda_R s_{ia}^{\ell}s_{ia}^{\ell+1}$ is absent when $\ell=L-1$.

We need to compute the expectation over the gaussian fields $h_a$.  Let us denote by $\cP_{q}(\{v_a,s_a\})$ the
probability that the Gaussian variables $h_1,...,h_n$ with mean zero
and covariance matrix $q$ satisfy the constraints $\forall a :\ s_a h_a+v_a>0$. Then
\begin{align}
    \overline{{\cal N}_\mathrm{fp}^n} &=  \int \prod_{a<b} dq_{ab} d\hat q_{ab}
    e^{-N\sum_\ell \sum_{a<b} \hat q^\ell_{ab}q^\ell_{ab}}\nonumber\\
  &\prod_{i} \left[\sum_{\{s^\ell_a\}}e^{ \sum_{\ell,a<b}\; \hat q^\ell_{ab}s_{a}^\ell s_{b}^\ell }\; 
  \prod_{\ell=1}^{L-1} \cP_{q^\ell+\Delta \pi_n}(\{J_D+\lambda_L s_{a}^{\ell-1}s_{a}^{\ell}+\lambda_R s_{a}^{\ell}s_{a}^{\ell+1},s_a^\ell\})
  \right] 
\end{align}

We now look at a replica symmetric Ansatz for the overlap matrices $q^\ell_{ab}$, defined by
\begin{align}
    q^\ell_{aa}=1\ \ ; \ \ q^\ell_{ab}=q^\ell\ \  \text{for}\ \ a\neq b
\end{align}
and same for $\hat q$.

 Given $q^\ell=q$, we can write the joint density of $\tilde h_1,...,\tilde h_n$ as
  \begin{align}
      P(\tilde h_1,...,\tilde h_n)=\int \frac{dz}{\sqrt{2\pi q}}e^{-z^2/(2q)}\ \prod_{a=1}^n \left[\frac{1}{\sqrt{2\pi(1-q)}}e^{-(\tilde h_a-z)^2/(2(1-q))}\right]
  \end{align}
  Therefore the function $\cP_q(\{v_a,s_a\})$  can be easily computed:
  \begin{align}
     {\cal P}_q(\{v_a,s_a\})=\int d  h_1...d  h_n\;  P( h_1,...,h_n)\; 
      \prod_{a=1}^n \theta( h_a s_a +v_a)
      = \int Dz\; \prod_{a=1}^n H\left(\frac{-v_a-s_{a} z\sqrt{q} }{\sqrt{1-q}}\right)
  \end{align}
In our case we have, in layer $\ell$, $v_a^\ell=J_D+\lambda_L s_{a}^{\ell-1}s_{a}^{\ell}+\lambda_R
s_{a}^{\ell}s_{a}^{\ell+1}$. 

Using 
\begin{align}
 e^{ \sum_{a<b}\; \hat q^\ell_{ab}s_{a}^\ell s_{b}^\ell }
 = e^{-n\hat q^\ell/2}\int Dt^\ell\;  e^{\sqrt{\hat q^\ell} t^\ell \sum_a s_a^\ell}\; 
\end{align}
we get
\begin{align}
    \overline{{\cal N}_\mathrm{fp}^n} &= \text{Extr}_{\{\hat q^\ell,q^\ell\}} 
    e^{\frac{Nn}{2}\sum_{\ell} \hat q^\ell (q^\ell-1)}\nonumber\\
  &\prod_{i} \left[
   \int \prod_{\ell=1}^{L} Dt^\ell Dz^\ell 
  \left(
  \sum_{s^1,...,s^{L}}
e^{\sum_\ell \sqrt{\hat q^\ell} t^\ell s^\ell}
\prod_{\ell} H\left(
\frac{-J_D-\lambda_L s^{\ell-1} s^\ell -\lambda_R s^{\ell} s^{\ell+1}-s^\ell z^\ell \sqrt{q^\ell}}{\sqrt{1-q^\ell}}
\right)
  \right)^n
  \right] 
  \label{Entropy_RS_general}
\end{align}
This gives finally
\begin{align}
    \frac{1}{NL} {\log \overline{{\cal N}^n}}=\frac{1}{L}\text{Extr}_{\{\hat q^\ell,q^\ell\}} 
    \left[\frac{1}{2}\sum_{\ell} \hat q^\ell (q^\ell-1)+\varphi(\{\hat q^\ell,q^\ell\})
  \right]
  \label{eq:result_RS}
\end{align}
where 
\begin{align}
    \varphi(\{\hat q^\ell,q^\ell\})= 
    \int \prod_{\ell=1}^{L} Dt^\ell Dz^\ell
    \log\left(
\zeta(\{\hat q^\ell,q^\ell,z^\ell,t^\ell\})
    \right)
\end{align}
  and 
\begin{align}
    \zeta(\{\hat q^\ell,q^\ell,z^\ell,t^\ell\})= 
    \sum_{s^1,...,s^{L}}
e^{\sum_\ell \sqrt{\hat q^\ell} t^\ell  s^\ell}
\prod_{\ell=1}^{L-1} H\left(
\frac{-J_D-\lambda_L s^{\ell-1} s^\ell -\lambda_R s^{\ell} s^{\ell+1}-s^\ell z^\ell \sqrt{q^\ell}}{\sqrt{1-q^\ell}}
  \right)
\end{align}

  We see that $\zeta $ is a partition functions of an Ising chain with
  next-nearest neighbor interactions, with gaussian external fields
  $t^\ell$ and $z^\ell$, and the quantities $\langle s^\ell\rangle^2$
  are the Edwards Anderson order parameter for the spins in layer
  $\ell$. So all this can be studied numerically adapting the transfer
  matrix already introduced in computation of the annealed average. Let us focus here on the cases $L=1,2$ which can be handled directly. We get:

   Case L=1:
 \begin{align}
   \zeta(\{\hat q^1,q^1,z^1,t^1\}) =
   e^{\sqrt{\hat q^{1}} t^{1} }
   H \left( \frac{-J_D-z^1
   \sqrt{q^1+\Delta}}{\sqrt{1-q^1}}\right)
   +
   e^{-\sqrt{\hat q^{1}} t^{1} }
   H \left(
   \frac{-J_D+z^1 \sqrt{q^1+\Delta}}
   {\sqrt{1-q^1}}
   \right) 
 \end{align}
 
Case L=2:
 \begin{align}
   \zeta &(\{\hat q^1,q^1,z^1,t^1,\hat q^2,q^2,z^2,t^2 \}) =\nonumber\\
   &e^{-\sqrt{\hat q^{1}} t^{1} -\sqrt{\hat q^{2}} t^{2}}
   H \left( \frac{-J_D
   -\lambda_R 
   +z^1  \sqrt{q^1+\Delta}}{\sqrt{1-q^1}}\right)
    H \left( \frac{-J_D
   -\lambda_L
   +z^2  \sqrt{q^2+\Delta}}{\sqrt{1-q^2}}
          \right)
    \nonumber \\                                                      
   &+
   e^{-\sqrt{\hat q^{1}} t^{1} +\sqrt{\hat q^{2}} t^{2}}
   H \left( \frac{-J_D
   /\lambda_R 
   +z^1  \sqrt{q^1+\Delta}}{\sqrt{1-q^1}}\right)
    H \left( \frac{-J_D
   +\lambda_L
   -z^2  \sqrt{q^2+\Delta}}{\sqrt{1-q^2}}
     \right)
     \nonumber \\                                                       
   &+
   e^{+\sqrt{\hat q^{1}} t^{1} -\sqrt{\hat q^{2}} t^{2}}
   H \left( \frac{-J_D
   +\lambda_R 
   -z^1  \sqrt{q^1+\Delta}}{\sqrt{1-q^1}}\right)
    H \left( \frac{-J_D
   +\lambda_L
   +z^2  \sqrt{q^2+\Delta}}{\sqrt{1-q^2}}
     \right)
       \nonumber \\                                                       
   &+
   e^{\sqrt{\hat q^{1}} t^{1} +\sqrt{\hat q^{2}} t^{2}}
   H \left( \frac{-J_D
   -\lambda_R 
   -z^1  \sqrt{q^1+\Delta}}{\sqrt{1-q^1}}\right)
    H \left( \frac{-J_D
   -\lambda_L
   -z^2  \sqrt{q^2+\Delta}}{\sqrt{1-q^2}}
      \right)
 \end{align}

 Using these expressions, we can study in detail the replica symmetric result for the number of fixed points.
 
 \textbf {Detailed RS study for a single core module, $L=1$}
 
The quenched evaluation with replica symmetric Ansatz gives:
\begin{align}
    S_{rs}=\frac{1}{N}\overline{\log \mathcal{N}}=\text{Extr}_{\{\hat q,q\}} \left[\frac{1}{2} \hat q(q-1)+
    \int Dt Dz\log\zeta(\hat q, q,z,t)
  \right]
\end{align}
where
\begin{align}
   \zeta(\{\hat q,q,z,t\}) =
   e^{\sqrt{\hat q^{}} t^{} }
   H \left( \frac{-J_D-z
   \sqrt{q}}{\sqrt{1-q}}\right)
   +
   e^{-\sqrt{\hat q^{}} t^{} }
   H \left(
   \frac{-J_D+z \sqrt{q}}
   {\sqrt{1-q}}
   \right) 
 \end{align}
This has a saddle point at $q=\hat q=0$, for which $\zeta= 2H(-J_D)$, giving back the annealed average.

\begin{figure}[t]
\centering
   \includegraphics[width=.6\textwidth]
  {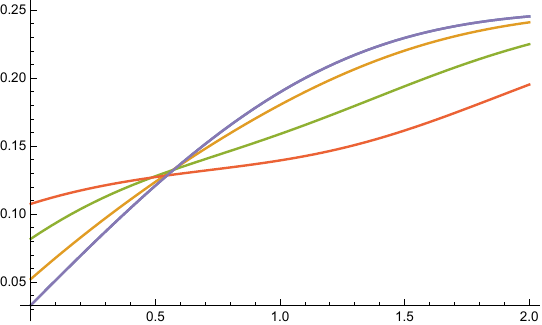}
  \caption{
  In the case with two core modules, $L=2$, we plot the coefficient of $q^2$  in the quadratic expansion of the RS free entropy. This coefficient is plotted versus $J_D$ for $\lambda=0,0.5,1,1.5$, from top to bottom on the right part of the figure (at $J_D=2$). The top blue curve with $\lambda=0$ is the same result as for $L=1$. The coefficient is positive for all values of $J_D,\lambda$ of interest. This shows that the RS saddle point at $q=\hat q=0$ is locally stable.
  }
  \label{fig.stab}
  \end{figure}

In order to check the stability of this saddle point, let us expand $S_{rs}$ to second order in powers of $q,\hat q$. We obtain:
\begin{align}
    S_{rs}^{(2)}=&\frac{1}{2} \hat q(q-1)+\log[2H(-J_D)]+\frac{1}{2}\hat q\nonumber\\
    &-\frac{1}{4}\hat q^2
    -\hat q q \frac{e^{-J_D^2}}
{4 \pi H(-J_D)^2}
-q^2\frac{J_D^2 e^{-J_D^2}}
{8 \pi H(-J_D)^2}
\end{align}
This is a concave function of $\hat q$, with a minimum at \begin{align}\hat q= q \frac{-2e^{-J_D^2}+4\pi H(-J_D)^2 }{4\pi H(-J_D)^2}
\end{align}
Evaluating $S_{rs}^{(2)}$ at this value of $\hat q$ we obtain
\begin{align}
    S_{rs}^{(2)}=\log[2H(-J_D)]+\left( 
    \frac{1}{4}+ 
    \frac{e^{-2 J_D^2}}{16\pi^2 H(-J_D)^4}\right)-
    \frac{(2+J_D^2)e^{-J_D^2}}{8\pi H(-J_D)^2}
\end{align}
which is a convex function of $q$ with minimum at $q=0$, for all $J_D\geq -.178$ (See Fig.\ref{fig.stab}). This shows that the saddle point at $q=\hat q=0$ is stable. We have not found any other saddle point. This confirms that the annealed average expression for the entropy of fixed points is exact.

  \textbf{Detailed RS study for $L=2$}
  
Taking $\Delta_V=\Delta_W=0$, we had found the annealed result:
\begin{align}
\frac{1}{2N}\log \overline {\cal N_\mathrm{fp}}=& \frac{1}{2} \log 2  + \frac{1}{2}\log \left[   H\left(-J_D-\lambda_R \right)H\left(-J_D -\lambda_L\right) + \right. \nonumber \\
     &      \left. H\left(-J_D+\lambda_R \right)
     H\left(-J_D +\lambda_L\right)
     \right]
\end{align}

Assuming that by symmetry the order parameters $q,\hat q$ are the same in each of the two layers, the quenched evaluation with replica symmetric Ansatz gives (see \ref{Entropy_RS_general}):

\begin{align}
    S_{rs}=\frac{1}{2N}\overline{\log \mathcal{N}}=\frac{1}{2}\text{Extr}_{\{\hat q,q\}} \left[\hat q(q-1)+
    \int Dt_1 Dz_1 D t_2 Dz_2 \log\zeta(\hat q, q,z_1,t_1,z_2,t_2)
  \right]
\end{align}
where
\begin{align}
   \zeta (\{\hat q,q,z_1,t_1,z_2,t_2 \}) &=\nonumber\\
   &e^{\sqrt{\hat q }(-t_{1} -t_{2})}
   H \left( \frac{-J_D
   -\lambda_R 
   +z_1  \sqrt{q}}{\sqrt{1-q}}\right)
    H \left( \frac{-J_D
   -\lambda_L
   +z_2  \sqrt{q}}{\sqrt{1-q}}
          \right)
    \nonumber \\ 
    &+ e^{\sqrt{\hat q }(-t_{1} +t_{2})}
   H \left( \frac{-J_D
   +\lambda_R 
   +z_1  \sqrt{q}}{\sqrt{1-q}}\right)
    H \left( \frac{-J_D
   +\lambda_L
   -z_2  \sqrt{q}}{\sqrt{1-q}}
          \right)   
     \nonumber \\
&+ e^{\sqrt{\hat q }(t_{1} -t_{2})}
   H \left( \frac{-J_D
   +\lambda_R 
   -z_1  \sqrt{q}}{\sqrt{1-q}}\right)
    H \left( \frac{-J_D
   +\lambda_L
   +z_2  \sqrt{q}}{\sqrt{1-q}}
          \right)   
     \nonumber \\   
     &+ e^{\sqrt{\hat q }(t_{1} +t_{2})}
   H \left( \frac{-J_D
   -\lambda_R 
   -z_1  \sqrt{q}}{\sqrt{1-q}}\right)
    H \left( \frac{-J_D
   -\lambda_L
   -z_2  \sqrt{q}}{\sqrt{1-q}}
          \right)   
 \end{align}

 In order to check the stability of the saddle point at $q=\hat q=0$, we proceed as for the case $L=2$: we  expand $S_{rs}$ to second order in powers of $q,\hat q$, find the saddle point value of $\hat q$ and substitute it into the expansion. Specializing to the case $\lambda_L=\lambda_R=\lambda$, the resulting quadratic form in $q$ is 
given by
\begin{align}
S_{rs}^{(2)}&=
\frac{1}{2}\left[\log 2 +\log(H(-J_D-\lambda)^2+H(-J_D+\lambda)^2)
\right]
\nonumber\\
+\frac{A(J_D,\lambda)}{B(J_D,\lambda)} \; q^2
\end{align}
where 
\begin{align}
   A(J_D,\lambda)= &2 \pi ^2 H(-J_D-\lambda )^8+8 \pi ^2 H(\lambda -J_D)^2 H(-J_D-\lambda
   )^6-4 e^{-(J_D+\lambda )^2} \pi  H(-J_D-\lambda )^6
   \nonumber \\ &
   -2 e^{-(J_D+\lambda
   )^2} J_D^2 \pi  H(-J_D-\lambda )^6
   -2 e^{-(J_D+\lambda )^2} \lambda ^2
   \pi  H(-J_D-\lambda )^6
   \nonumber \\ &-4 e^{-(J_D+\lambda )^2} J_D \lambda  \pi 
   H(-J_D-\lambda )^6
   -4 e^{3 J_D^2+8 \lambda  J_D+3 \lambda ^2-4
   (J_D+\lambda )^2} J_D^2 \pi  H(\lambda -J_D) H(-J_D-\lambda )^5
   \nonumber \\ &+4 e^{3
   J_D^2+8 \lambda  J_D+3 \lambda ^2-4 (J_D+\lambda )^2} \lambda ^2 \pi 
   H(\lambda -J_D) H(-J_D-\lambda )^5
   +e^{-2 (J_D+\lambda )^2} H(-J_D-\lambda
   )^4
   \nonumber \\ &
   +2 e^{4 J_D \lambda -2 (J_D+\lambda )^2} H(-J_D-\lambda )^4-e^{8 J_D
   \lambda -2 (J_D+\lambda )^2} H(-J_D-\lambda )^4+12 \pi ^2 H(\lambda
   -J_D)^4 H(-J_D-\lambda )^4
   \nonumber \\ &
   -8 e^{-(J_D+\lambda )^2} \pi  H(\lambda -J_D)^2
   H(-J_D-\lambda )^4-4 e^{4 J_D \lambda -(J_D+\lambda )^2} \pi  H(\lambda
   -J_D)^2 H(-J_D-\lambda )^4
   \nonumber \\ &
   -2 e^{4 J_D \lambda -(J_D+\lambda )^2} J_D^2 \pi 
   H(\lambda -J_D)^2 H(-J_D-\lambda )^4
   -2 e^{4 J_D \lambda -(J_D+\lambda
   )^2} \lambda ^2 \pi  H(\lambda -J_D)^2 H(-J_D-\lambda )^4
   \nonumber \\ &
   +4 e^{4 J_D
   \lambda -(J_D+\lambda )^2} J_D \lambda  \pi  H(\lambda -J_D)^2
   H(-J_D-\lambda )^4
   +8 \pi ^2 H(\lambda -J_D)^6 H(-J_D-\lambda )^2
   \nonumber \\ &-4
   e^{-(J_D+\lambda )^2} \pi  H(\lambda -J_D)^4 H(-J_D-\lambda )^2-8 e^{4
   J_D \lambda -(J_D+\lambda )^2} \pi  H(\lambda -J_D)^4 H(-J_D-\lambda
   )^2
   \nonumber \\ &
  -2 e^{-(J_D+\lambda )^2} J_D^2 \pi  H(\lambda -J_D)^4 H(-J_D-\lambda
   )^2
   -2 e^{-(J_D+\lambda )^2} \lambda ^2 \pi  H(\lambda -J_D)^4
   H(-J_D-\lambda )^2
   \nonumber \\ &
   -4 e^{-(J_D+\lambda )^2} J_D \lambda  \pi  H(\lambda
   -J_D)^4 H(-J_D-\lambda )^2+4 e^{2 \left(J_D^2+4 \lambda  J_D+\lambda
   ^2\right)-4 (J_D+\lambda )^2} H(\lambda -J_D)^2 H(-J_D-\lambda )^2
   \nonumber \\ &-4
   e^{3 J_D^2+8 \lambda  J_D+3 \lambda ^2-4 (J_D+\lambda )^2} J_D^2 \pi 
   H(\lambda -J_D)^5 H(-J_D-\lambda )
   \nonumber \\ &
   +4 e^{3 J_D^2+8 \lambda  J_D+3 \lambda
   ^2-4 (J_D+\lambda )^2} \lambda ^2 \pi  H(\lambda -J_D)^5 H(-J_D-\lambda
   )+2 \pi ^2 H(\lambda -J_D)^8
   \nonumber \\ &
   -4 e^{J_D^2+6 \lambda  J_D+\lambda ^2-2
   (J_D+\lambda )^2} \pi  H(\lambda -J_D)^6-2 e^{J_D^2+6 \lambda 
   J_D+\lambda ^2-2 (J_D+\lambda )^2} J_D^2 \pi  H(\lambda -J_D)^6
   \nonumber \\ &
   -2
   e^{J_D^2+6 \lambda  J_D+\lambda ^2-2 (J_D+\lambda )^2} \lambda ^2 \pi 
   H(\lambda -J_D)^6
   +4 e^{J_D^2+6 \lambda  J_D+\lambda ^2-2 (J_D+\lambda
   )^2} J_D \lambda  \pi  H(\lambda -J_D)^6
   \nonumber \\ &
   -e^{-2 (J_D+\lambda )^2}
   H(\lambda -J_D)^4+2 e^{4 J_D \lambda -2 (J_D+\lambda )^2} H(\lambda
   -J_D)^4+e^{8 J_D \lambda -2 (J_D+\lambda )^2} H(\lambda -J_D)^4
\end{align}
and
\begin{align}
   B(J_D,\lambda)= 16 \pi ^2 \left(H(-J_D-\lambda)^2+H(\lambda-J_D)^2\right)^2
   \left(H(-J_D-\lambda)^4+H(\lambda-J_D)^4\right)
\end{align}
Fig. \ref{fig.stab} shows that the ratio $A(J_D,\lambda)/B(J_D,\lambda)$ is always positive in the ranges of $J_D,\lambda$ of interest. This shows that the saddle point at $q=\hat q=0$ is locally stable, which indicates that the annealed result for the entropy of fixed points is correct also for this case with $L=2$.

%% file: local_entropy.tex
Here we derive the expression for the free-energy density $\Phi(d, y)$ within the local entropy approach. We  follow  ref. [13] 
and derive the local entropy theory through a 1-RSB formalism.
We  want to compute the number of fixed points  $\mathcal{N}(\tilde s,d)$  at distance $d$ from an optimal reference configuration $\tilde s$. We adopt the short notation $\cal N$ for ${\cal N}({\tilde s}, y)$ and $\Phi$  for $\Phi(d, y)$.
We  study the quenched average of $\log \mathcal{N}$ with respect to the choice the internal couplings in each module, using $n$ replicas, for a generic multilayer model consisting of $L$ layers of core modules.  The $n$-th power of $\mathcal N $ is expressed as:
\begin{align}
    {\cal N}^n = \sum_{s} \prod_{\ell=1}^{L} \prod_{i=1}^N \prod_{a=1}^n \theta\left(  s_{ia}^\ell h_{i}^\ell +J_D+\lambda s_{ia}^{\ell-1}s_{ia}^{\ell}+\lambda s_{ia}^{\ell}s_{ia}^{\ell+1}   \right)
\end{align}
where the index $\ell\in\{1,...,L\}$ denotes the layer, the index $i\in\{1,...,N\}$ denotes the neuron within each layer, the index $a\in\{1,...,n\}$ is a replica index, the $\sum_{s_a}$ denotes a sum over all the $2^{L N n}$ internal
representations of the variables $ s_{ia}^\ell $, and $\theta $ is Heavyside's step function.

When sampling the $J$ internal to each module to perform the quenched average, the local fields $ h_{ia}^{\ell }=  \sum_j J_{ij}^\ell s_{ja}^{\ell}$
have a joint Gaussian distribution with mean zero. They are independent for each $i,\ell$ and their covariance in replica space is 
\begin{align}
\mathbb{E} \; h_{ia}^{\ell } h_{jb}^{m  }= \delta_{i j}\delta^{\ell m} q^\ell_{ab}
\end{align}
where
\begin{align}
    q^\ell_{ab}=\frac{1}{N}\sum_j s_{ja}^{\ell }s_{jb}^{\ell }
\end{align}

Introducing this replica overlap $ q^\ell_{ab}$ in each layer $\ell$ and the corresponding Lagrange multiplier  $\hat  q^\ell_{ab}$, we find the expression:
\begin{align}
    \overline{{\cal N}^n }&= \sum_{s} \int \prod_{l=1}^L \prod_{a<b} \frac{dq_{ab}^\ell d\hat q_{ab}^\ell}{2\pi}
    e^{-N\sum_\ell \sum_{a<b} \hat q^\ell_{ab}q^\ell_{ab}+\sum_{\ell,i,a,b}\hat q^\ell_{ab}s_{ia}^\ell s_{ib}^\ell }\nonumber\\
  &\prod_{i\ell} \left[
    \mathbb{E}_{\{h_a\}\sim{\cal N}(0,q^\ell _{ab})} \;\prod_{a=1}^n \theta\left(s_{ia}^\ell h_{a}+J_D+\lambda s_{ia}^{\ell-1}s_{ia}^{\ell}+\lambda s_{ia}^{\ell}s_{ia}^{\ell+1}\right)
  \right] 
  \label{Nnav}
\end{align}
where the integrals over the $\hat q$ variables run along the imaginary axis.

Let us denote by $\cP_{q}(\{v_a,s_a\})$ the
probability that the Gaussian variables $h_1,...,h_n$ with mean zero
and covariance matrix $q$  satisfy the constraints $\forall a :\ s_a h_a+v_a>0$. This is the quantity which appears in the last line of (\ref{Nnav}).

Let us assume that each $q^\ell$ matrix has a 1-RSB structure [25, 26] 
with diagonal equal to $1$, elements $q_1$ inside blocks of size $m$ and elements $q_0$ outside of the diagonal blocks. Then we can compute the explicit form of $\cP_{q}(\{v_a,s_a\})$. We first write the joint density of $h_1,..., h_n$ as:
  \begin{align}
      &P(h_1,..., h_n)=\int \frac{dz_0}{\sqrt{2\pi (q_0)}}e^{-z_0^2/(2 q_0)}\nonumber\\
      &\prod_{B=1}^{n/m}\left(\int \frac{dz_B}{\sqrt{2\pi (q_1-q_0)}}
      e^{-(z_B-z_0)^2/(2(q_1-q_0))}\ \prod_{a\in B} \left[\frac{1}{\sqrt{2\pi(1-q_1)}}e^{-(h_a-z_B)^2/(2(1-q_1))}\right]
      \right)
  \end{align}
  where the index $B$ runs over the various blocks of the 1-RSB Ansatz, and in each block the index $a$ can take $m$ values.
From this expression  we deduce that, when $q$ is a 1-RSB matrix with parameters $q_0,q_1,1$, one has:
  \begin{align}
      \cP_q(\{v_a,s_a\})=\int D_{q_0}(z_0)
      \prod_{B=1}^{n/m}\left(\int D_{q_1-q_0}(z_B-z_0) \prod_{a\in B} H\left(\frac{-v_a-z_B s_a}{\sqrt{1-q_1}}\right)
      \right)\ ,
  \end{align}
where we use  the notation:
  \begin{align}
      D_q(z)=\frac{dz}{\sqrt{2\pi q}}\ e^{-z^2/(2q)} \ \  ;\ \ H(x)=\int_x^\infty D_1(z)=\frac{1}{2}
      \text{Erfc}
      \left(\frac{x}{\sqrt{2}}\right)
\end{align}
Then we have 
\begin{align}
    \overline{{\cal N}^n} &=  \int \prod_{a<b} dq^\ell_{ab} d\hat q^\ell_{ab}
    e^{-N\sum_\ell \sum_{a<b} \hat q^\ell_{ab}q^\ell_{ab}}Z(q,\hat q)^N
    \end{align}
where
\begin{align}
 Z(q,\hat q)&=\sum_{\{s^\ell_a\}}e^{ \sum_{\ell,a<b}\; \hat q^\ell_{ab}s_{a}^\ell s_{b}^\ell }\; 
  \prod_{\ell=1}^{L} {\cal P}_{q^\ell}(\{J_D+\lambda s_{a}^{\ell-1}s_{a}^{\ell}+\lambda s_{a}^{\ell}s_{a}^{\ell+1},s_a^\ell\}) \\
  &= \sum_{\{s^\ell_a\}}e^{ \sum_{\ell,a<b}\; \hat q^\ell_{ab}s_{a}^\ell s_{b}^\ell }\; 
  \prod_{\ell=1}^{L} \int D_{q_0^\ell}(z_0^\ell)
      \prod_{B}\left(\int D_{q_1^\ell-q_0^\ell}(z_B^\ell-z_0^\ell) \right. \nonumber \\ &\left. \prod_{a\in B} H\left(\frac{-(J_D+\lambda s_a^{\ell-1}s_a^\ell +\lambda s_a^{\ell+1}s_a^\ell+z_B^\ell  s_a^\ell)}{\sqrt{1-q_1^\ell}}\right)
      \right)
\end{align}

When the $\hat q^\ell$ matrices have a 1-RSB structure with  elements $\hat q_1^\ell$ inside blocks of size $m$ and elements $\hat q_0$ outside of the diagonal blocks, we can write 
\begin{align}
 e^{ \sum_{a<b}\; \hat q_{ab}^\ell s_{a}^\ell s_{b}^\ell }
 =e^{-n \hat q_1/2}\int D_{\hat q_0^\ell}(t_0^\ell)\prod_B\left[
 \int D_{\hat q_1^\ell-\hat q_0^\ell}(t_B^\ell-t_0^\ell) e^{t_B^\ell\sum_{a\in B} s_a}
 \right]
\end{align}
so that
\begin{align}
 Z(q,\hat q)&=
  \prod_{\ell=1}^{L}\left[ \int D_{q_0^\ell}(z_0^\ell) \int D_{\hat q_0^\ell}(t_0^\ell)
   \prod_{B}\left(\int D_{q_1^\ell-q_0^\ell}(z_B^\ell-z_0^\ell) \int D_{\hat q_1^\ell -\hat q_0^\ell}(t_B^\ell-t_0^\ell) \right)\right]\nonumber \\
   &e^{-(n/2)\sum_\ell \hat q_1^\ell} \sum_{\{s^\ell_a\}} e^{\sum_{\ell,a} s_a^\ell t_{B(a)}^\ell}
   \prod_{\ell,a} H\left(\frac{-(J_D+\lambda_L s_a^{\ell-1}s_a^\ell +\lambda_R s_a^{\ell+1}s_a^\ell+z_B^\ell  s_a^\ell)}{\sqrt{1-q_1^\ell}}\right)
   \label{Zdef_general}
\end{align}

We now analyze the core module and the two layer chain.

\subsubsection{Core module}
 Let us specialize first to the case $L=1$, which corresponds to the bare core module.
 We find:
 \begin{align}
 Z(q,\hat q)&=
  \int D_{q_0}(z_0) \int D_{\hat q_0}(t_0)
   \prod_{B}\left(\int D_{q_1-q_0}(z_B-z_0) \int D_{\hat q_1 -\hat q_0}(t_B-t_0) \right)\nonumber \\
   &e^{-n\hat q_1/2}\sum_{\{s_a\}} e^{\sum_{a} s_a t_{B(a)}}
   \prod_{a} H\left(
   -\frac{J_D+z_{B(a)}  s_a}{\sqrt{1-q_1}}\right)\\
   \end{align}
   and therefore, in the $n\to 0$ limit:
   \begin{align}
   \log Z=&-\frac{n}{2}\hat q_1+\frac{n}{m}
    \int D_{q_0}(z_0) \int D_{\hat q_0}(t_0)
  \log \left(\int D_{q_1-q_0}(z_1-z_0) \int D_{\hat q_1 -\hat q_0}(t_1-t_0) \right.\nonumber \\
   &\left.
  \left[
   e^{t_1} H\left(
   -\frac{J_D+z_1}{\sqrt{1-q_1}}\right) + e^{-t_1}H\left(
   -\frac{J_D-z_1}{\sqrt{1-q_1}}\right)\right]^m\right)
 \end{align}
The final result for the free-energy density at fixed $q_1= 1-2 d$ and $m=y$ is 
\begin{align}
    -y \Phi(y)=\frac{1}{N}\overline {\log \mathcal N} &=\text{Extr}_{\hat q_1,\hat q_0,q_0}\left[
    -\frac{1}{2}\hat q_1+
    \frac{y}{2}\hat q_0 q_0 +\frac{1-y}{2}\hat q_1 q_1 
    \right]\nonumber\\
    &+\frac{1}{y}\int D_{q_0}(z_0) \int D_{\hat q_0}(t_0)
  \log \left(\int D_{q_1-q_0}(z_1-z_0) \int D_{\hat q_1 -\hat q_0}(t_1-t_0) \right.\nonumber \\
   &\left.
  \left[
   e^{t_1} H\left(
   -\frac{J_D+z_1}{\sqrt{1-q_1}}\right) + e^{-t_1}H\left(
   -\frac{J_D-z_1}{\sqrt{1-q_1}}\right)\right]^y\right)
 \end{align}
It can be rewritten as 
\begin{align}
    -y\Phi(y)&=\text{Extr}_{\hat q_1,\hat q_0,q_0}\left[
    -\frac{1}{2}(\hat q_1-\hat q_0+\hat q_0)+
    \frac{y}{2}\hat q_0 q_0 \right. \nonumber \\
    &\left.+\frac{1-y}{2}\left((\hat q_1 -\hat q_0)(q_1-q_0)+(\hat q_1 -\hat q_0)q_0+\hat q_0 (q_1-q_0)+\hat q_0 q_0\right)
    \right]\nonumber\\
    &+\frac{1}{y}\int Dz_0 Dt_0 
  \log \left(\int Dz_1 Dt_1 
  \left[\sum_s
   e^{s X } H\left(
   -\frac{J_D+s Y }{\sqrt{1-q_1}}\right) \right]^y\right)
 \end{align}
 where:
 \begin{align}
     X=&t_0\sqrt{\hat q_0}+t_1\sqrt{\hat q_1-\hat q_0}\\
     Y=&z_0\sqrt{q_0}+z_1\sqrt{q_1 -q_0}
 \end{align}

We need to compute the saddle point over $q_0$, $\hat q_0$ and $\hat q_1$. 
 It is clear that the stationnarity equations with respect to $q_0$ and $\hat q_0$ have the solution $q_0=\hat q_0=0$ which correspond to the symmetry of our measure by flipping simultaneously all the fields. 
 We shall assume that this symmetry is not broken. Then we get:
\begin{align}
   -y \Phi(y)=
    -\frac{1}{2}\hat q_1+\frac{1-y}{2}\hat q_1 q_1 
   +
    \frac{1}{y}
  \log \left(\int Dz Dt 
  \left[\sum_s
   e^{s t\sqrt{\hat q_1} } H\left(
   -\frac{J_D+s z\sqrt{q_1} }{\sqrt{1-q_1}}\right) \right]^y\right)
 \end{align}
 where $\hat q_1$ is the solution of the stationnarity equation
\begin{align}
     q_1&= \frac{
     \int Dz Dt
     \left[\sum_s\; 
   s \; e^{s t\sqrt{\hat q_1}} H\left(
   -\frac{J_D+s z\sqrt{q_1} }{\sqrt{1-q_1}}\right) \right]^2
  \left[\sum_s
   e^{s t\sqrt{\hat q_1}} H\left(
   -\frac{J_D+s z\sqrt{q_1} }{\sqrt{1-q_1}}\right) \right]^{y-2}
     }{
     \int Dz Dt 
  \left[\sum_s
   e^{s t\sqrt{\hat q_1} } H\left(
   -\frac{J_D+s z\sqrt{q_1} }{\sqrt{1-q_1}}\right) \right]^y
     }
 \end{align}

\subsubsection{Two-layer chain}
We go back to the general expression (\ref{Zdef_general}) and specialize to the case $L=2$. Using the symmetry between the two layers, it is reasonable to assume that 
$q_0^1=q_0^2=q_0$, $q_1^1=q_1^2=q_1$ and similarly for the $\hat q$ elements. At the saddle point $q_0=\hat q_0=0$, we get 
\
 \begin{align}
    -y \Phi(d,y)=& 
    -\hat q_1
    +(1-y) \hat q_1 q_1 
  +\frac{1}{y}\
  \log \left(\int Dz Dz' Dt Dt' \right.\nonumber \\
   & \left.\left[
   \sum_{s,s'}
   e^{\sqrt{\hat q_1} (s t+s' t')}
 H\left(
   \frac{-J_D+\lambda s s'-s\sqrt{q_1} z}{\sqrt{1-q_1}}\right)
   H\left(
   \frac{-J_D+\lambda s s'-s'\sqrt{q_1} z'}{\sqrt{1-q_1}}
   \right)
   \right]^m
   \right)
 \end{align}
Where $\hat q_1$ is the solution of the stationarity equation:

     \begin{align}
     q_1&= \frac{
     \int Dz Dz' Dt Dt' 
     \left[\sum_{s,s'}\; s\; 
   e^{\sqrt{\hat q_1} (s t+s' t')} H^{(2)}_{s s'} \right]^2
  \left[\sum_{s,s'}
   e^{\sqrt{\hat q_1} (s t+s' t')} H^{(2)}_{s s'} \right]^{m-2}
     }{
     \int Dz Dz' Dt Dt'  
  \left[\sum_{s,s'} 
   e^{\sqrt{\hat q_1} (s t+s' t')} H^{(2)}_{s s'} \right]^m
     }
 \end{align}
 where 
 \begin{align}
     H^{(2)}_{s s'}=  H\left(
   \frac{-J_D+\lambda s s'-s\sqrt{q_1} z}{\sqrt{1-q_1}}\right)
   H\left(
   \frac{-J_D+\lambda s s'-s'\sqrt{q_1} z'}{\sqrt{1-q_1}}
   \right)
 \end{align}

%% file: training-variants.tex
\label{sec:training-variants}
In this section, we discuss some extensions of the training procedure described in the main text, summarized at a high level in Algorithm \ref{alg:supervised_learning}.

\begin{algorithm}[ht]
\caption{General Supervised Learning Protocol for Recurrent Neural Networks}
\label{alg:supervised_learning}
\Input{Training set $\{(x^\mu, y^\mu)\}_{\mu=1}^P$; initial weights $J\in\mathbb{R}^{N\times N}$, $W^{in}\in\mathbb{R}^{N\times D}$, $W^{out}\in\mathbb{R}^{C\times N}$; class prototypes $W^{back}\in\mathbb{R}^{N\times C}$; adjacency $A\in\{0,1\}^{N\times N}$; binary vectors $a^{in},a^{back},a^{out}\in\{0,1\}^N$; learning rate $\eta$; margin $\kappa$; external field strengths $\boldsymbol{\lambda}=(\lambda_x,\lambda_y)$.}
\Output{Updated synaptic weights $J, W^{in}$; updated readout weights $W^{out}$.}
$\boldsymbol{\lambda}^{free} \gets (\lambda_x, 0)$\;
\For{$\mu \gets 1$ \KwTo $P$}{
  $\;\;\;\; s \gets 0$\;
  $\;\;\;\; s \gets \mathrm{Warmup}\bigl(J, s, x^\mu, \boldsymbol{\lambda}^{free} \bigr)$ \tcp*{Synchronous evolution (Eq.~\ref{eq:general-topology-update})}
  $\;\;\;\; s \gets \mathrm{EvolveUntilConvergenceSupervised}\bigl(J, s, x^\mu, \boldsymbol{\lambda}, y^\mu \bigr)$ \tcp*{Synchronous evolution (Eq.~\ref{eq:general-topology-update})}
  $\;\;\;\; s \gets \mathrm{EvolveUntilConvergenceFree}\bigl(J, s, x^\mu, \boldsymbol{\lambda}^{free} \bigr)$ \tcp*{Synchronous evolution (Eq.~\ref{eq:general-topology-update})}
  $\;\;\;\; J, W^{in} \gets \mathrm{UpdateSynapticWeights}\bigl(J, W^{in}, s, x^\mu, \eta, \kappa\bigr)$ \tcp*{Synaptic plasticity (Eqs.~\ref{eq:hidden-field-general}--\ref{eq:plasticity_rule-win})}
  $\;\;\;\; W^{out} \gets \mathrm{UpdateReadoutWeights}\bigl(W^{out}, s, y^\mu, \eta, \kappa\bigr)$ \tcp*{Readout update (Eqs.~\ref{eq:hidden-field-wout}--\ref{eq:plasticity_rule-wout})}
}
\textbf{Inference:} Given input $x$, compute $\hat{s} \gets \mathrm{EvolveUntilConvergenceFree}\bigl(J, 0, x, \boldsymbol{\lambda}^{free} \bigr)$; output $\hat{y}=W^{out}\hat{s}$.
\end{algorithm}

\paragraph{\textbf{Networks with arbitrary topology}}

First, we generalize the learning algorithm, which was described for the core module in the main text, to recurrent networks with an arbitrary internal topology.

Assume that we are provided with a set of \(P\) patterns with their associated labels \( \{ x^\mu , y^\mu \}_{\mu=1}^P\) with \(x^\mu \in \mathbb{R}^D, y^\mu \in \mathbb{R}^C\). 
Consider a generic recurrent network of binary neurons \(s_1, \dots, s_N \in \{-1, 1\}\) that interact through a coupling matrix \(J \in \mathbb{R}^{N \times N}\). Assume that the topology of the network is encoded via an adjacency matrix \(A \in \{0, 1\}^{N \times N}\), such that \(A_{ij} = 0\) implies \(J_{ij} = 0\). Note that the core module and the multilayer chain model can also be represented this way by an appropriate choice of \(A\) and \(J\).

As for the core module, we introduce projection matrices \(W^{in} \in \mathbb{R}^{N \times D}\), \(W^{back} \in \mathbb{R}^{N \times C}\) and \(W^{out} \in \mathbb{R}^{C \times N}\). Not all neurons are necessarily affected by the input or the supervisory signal, nor do they all necessarily affect the predicted output. We encode this with binary vectors \(a^{in}, a^{back}, a^{out} \in \{0,1\}^N\).

In this more general case, the network evolves in time according to the update rule:
\begin{equation}
    s_i \leftarrow \sign \Big( \sum_{j=1}^N A_{ij} \, J_{ij} \, s_j + a^{in}_i \, \lambda_x \sum_{k=1}^D  W^{in}_k x^\mu_k + a^{back}_i \, \lambda_y \sum_{c=1}^C W^{back}_c y^\mu_c ) \Big) \;,
    \label{eq:general-topology-update}
\end{equation}
Like for the core module, when an equilibrium is reached, the supervisory signal is removed (set \(\lambda_y = 0\)), and the dynamics continues until a new fixed point is reached. Denote by \(s^\prime\) and \(s^{*}\) the two internal states reached at the end of the two phases. The plasticity rule in the general case can be written as:
\begin{align}
    \label{eq:hidden-field-general}
    h_i &= \sum_{j=1}^N A_{ij} \, J_{ij} \, s_j^* + a^{in}_i \, \lambda_x \sum_{k=1}^D  W^{in}_k x^\mu_k \\
    J_{ij} &\leftarrow J_{ij} + A_{ij} \, \eta \, s_i^{*} \, s_j^{*} \; \mathds{1}(s_i^{*} \cdot h_i \le k)
    \label{eq:plasticity_rule-general}
\end{align}
The projection \(W^{in}\) can be learned using the same rule, modifying Eq. \eqref{eq:plasticity_rule-general} appropriately: \(x^\mu_k\) plays the role of \(s_j^{*}\) and \(a^{in}_i\) plays that of \(A_{ij}\) to update each component \(W^{in}_{ik}\). In formulas:
\begin{align}
    \label{eq:plasticity_rule-win}
    W^{in}_{ik} &\leftarrow W^{in}_{ik} + a^{in}_i \, \eta \, s_i^{*} \, x^\mu_k \; \mathds{1}(s_i^{*} \cdot h_i \le k)
\end{align}
As for the readout matrix, many options are possible, depending on the specific task at hand. In the case of classification, assuming \(y^\mu\) is one-hot encoded using ones to signal the correct class and minus ones to signal the wrong ones, we can employ an analogous perceptron learning rule:
\begin{align}
    \label{eq:hidden-field-wout}
    l_c &= \sum_{j=1}^N a^{out}_j W^{out}_{cj} s^*_j \\
    W^{out}_{cj} &\leftarrow W^{out}_{cj} + a^{out}_j \, \eta \, y^\mu_c \, s_j^{*} \; \mathds{1}(y^\mu_c \cdot l_c \le k)
    \label{eq:plasticity_rule-wout}
\end{align}
Note that learning the readout is completely decoupled from the synaptic plasticity affecting \(J\) and \(W^{in}\). \(W^{out}\) could be left untrained until the very end of training, and then be learned offline, with completely equivalent results. Also note that, to accurately reduce this general case to the core module or the multilayer chain model, we need to avoid updating \(J_D\) and/or \(\lambda\).

In this general formulation, we can use \(A\) to encode the internal topology of the network. For example, we can generalize the multilayer chain model to allow nonzero couplings between any pair of neurons in subsequent layers (a 'fully-connected' multi-layer design). This corresponds to a 'block-tridiagonal' structure for \(A\): each 'block' encodes the interactions (or lack thereof) between a pair of layers; \(A\) has ones exactly in all entries of all blocks lying along the main diagonal, the first subdiagonal and the first supradiagonal. We can also use the binary vectors \(a^{in}, a^{back}, a^{out}\) to control the connectivity between the layers and the input/output. For example, we can introduce skip connections between intermediate layers and the input/output, or between distant layers.

We have experimented with these and other modifications to the network topology. Our preliminary results suggest that, provided the  RM is present, the system exhibits nontrivial learning capabilities using the algorithm described. This flexibility is promising, since it means that it could be possible to introduce an inductive bias via the network topology (e.g., exploiting the spatial structure of images via convolutions).

\paragraph{\textbf{Mini-batch training}} It is possible to group the data patterns into mini-batches of size $B$ to speed up the training process, instead of showing the training examples one by one. 
The network relaxation is carried out independently for each pair \((x^\mu, y^\mu)\) in the batch, in parallel, exploiting batched tensor operations. This provides fixed points \((s')^\mu\) and \((s^*)^\mu\) for each. Then, the couplings updates are computed independently and in parallel for each, and the individual updates are averaged along the batch dimension. In formulas, \eqref{eq:hidden-field-general} and \eqref{eq:plasticity_rule-general} become:
\begin{align}
    h_i^\mu &= \sum_{j=1}^N A_{ij} \, J_{ij} \, (s^*)_j^\mu + a^{in}_i \, \lambda_x \sum_{k=1}^D  W^{in}_k x^\mu_k , \quad\quad \mu = 1, \dots, B\\
    J_{ij} &\leftarrow J_{ij} + \frac{\eta}{B} \sum_{\mu=1}^B A_{ij} \, (s^{*})_i^\mu \, (s^{*})_j^\mu \; \mathds{1}((s^{*})_i^\mu \cdot h_i^\mu \le k)
\end{align}

\paragraph{\textbf{Multilayer chain model}}
We report here an observation about the synchronous update dynamics with the multilayer chain model.
The synchronous dynamics allows to exploit the parallelism of modern computing hardware like GPUs. However, especially in the multilayer chain model, this fully parallel update can lead to the onset of cycles. Intuitively, this is because the strong positive couplings between pairs of neurons belonging to subsequent layers tend to keep swapping their states, if they initially differ. To mitigate this effect, when using the multilayer chain model, we introduce a hyperparameter \(p \in (0, 1]\): at each time step, instead of updating all neurons at once according to \eqref{eq:multilayer-chain-update}, we update each neuron with probability \(p\).
This simple strategy, which preserves the distributed character of the dynamics and does not require global coordination, is effective at avoiding limit cycles.

\begin{figure}
    \centering
    \includegraphics[width=0.6\linewidth]{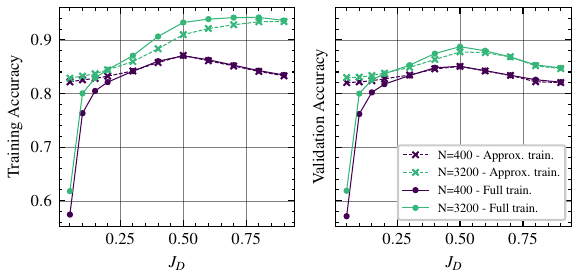}
    \caption{Comparison of full and approximate training dynamics on the Entangled-MNIST dataset with $N = 400$ and $N = 1600$ for $J_D \in [0.1, 0.9]$. Results are averaged over two independent runs, each trained for 100 epochs. The approximate scheme omits the unsupervised relaxation phase by setting $s^* := s'$.} 
    \label{fig:single-double}
\end{figure}

\paragraph{\textbf{Approximate training scheme}}  
A key feature of the proposed training protocol is that, once the initial \textit{supervised} fixed point \(s^\prime\) is reached and the teaching signal is removed, the network state remains effectively stable—only the local preactivation fields are altered. More formally, in Table \ref{tab:sprime_star}, we observe that:
\begin{equation}
    \frac{1}{N P} \sum_{\mu=1}^P \sum_{i=1}^N s^{\prime \mu}_i\, s^{*\mu}_i \approx 1,
\end{equation}
indicating that \(s^\prime\) and \(s^*\) are nearly identical across patterns. 
Motivated by this observation, we propose an \textit{approximated} training scheme where the final free phase of the dynamics is skipped (Line 6 in Algorithm \ref{alg:supervised_learning}), and the local plasticity rule is applied directly using the supervised fixed point.
Empirically, we find that this approximation preserves overall classification performance in networks with narrower hidden layers (\(N \sim 400\)), while wider networks (\(N \sim 3200\)) exhibit a measurable reduction in training and validation accuracies in the classification-optimal $J_D$ regime; see Figure~\ref{fig:single-double}. 

Notably, the approximate training scheme enables the network to learn input-output relations even at low self-coupling intensities (\(J_D \sim 0.1\)), which typically lead to unstable or chaotic dynamics in the full model. One contributing factor may be the absence of the separate unsupervised relaxation phase: in the approximate scheme, the system evolves under the simultaneous influence of both input and target fields (modulated by \(\lambda_x\) and \(\lambda_y\), respectively), in addition to the self-interaction term \(J_D\). This combined structure imposes stronger constraints on the dynamics and may help suppress chaotic behavior when \(J_D\) is weak. 

Note that, contrary to the algorithm presented in the main text, this variant effectively ignores the contribution of the term \(\lambda_y \;W^{back} y^\mu\) to the local field at the moment of the plasticity step. Such a 'special treatment' of the synapses providing supervision, which are used during the dynamics but not during the plasticity, is not necessary in the algorithm presented in the main text. This makes the approximate training scheme less appealing from a biological perspective.

\paragraph{\textbf{Local regularization}} Modern neural networks are typically trained by minimizing a task-specific loss function, which quantifies the discrepancy between predicted outputs and target labels. 
Beyond guiding learning, the choice of loss function implicitly shapes the geometry and robustness of the learned representations. 
For instance, the commonly used cross-entropy loss not only promotes correct classification but also encourages large decision margins between classes. 
This margin-based behavior has been linked to improved generalization and robustness, particularly in the presence of noise or adversarial perturbations. Other losses, such as hinge or contrastive losses, can be explicitly designed to enforce margin maximization or structural constraints on the embedding space. 
The proposed training protocol is not based on a global loss function, but can nonetheless accommodate for such kind of regularization by exploiting the local nature of its update rule.
Specifically, we can introduce a surrogate to cross entropy loss in the couplings update rule (Eq. \ref{eq:plasticity_rule}):
\begin{equation}
    J_{ij} \gets J_{ij} + \eta s^*_i s^*_j\left[-\frac{1}{2} + \frac{1}{2}\tanh\big(\gamma (s_i^* h_i - \kappa)\big)\right].
\end{equation}
Instead of simply increasing the stability margin conditionally on a threshold $\kappa$, we modulate synaptic updates based on a smooth function of $\kappa$ that encourages both \textit{correct} and \textit{confident} predictions.
In this way, the model inherits the generalization benefits of margin-based objectives, akin to those induced by cross-entropy, while remaining fully local and compatible with biologically plausible learning mechanisms.

%% file: benchmarks-hyperparameters-and-details.tex
In this section we provide full details on the numerical experiments to ensure full reproducibility of our results. Raw experimental data is also available on the Weights and Biases platform upon request to the authors.

\subsubsection{Recurrent model and Multi Layer Perceptron}
We first provide details on the first set of benchmarks, that compare our recurrent model against 2-hidden layer multilayer perceptron for fixed amounts of trainable parameters.
For the recurrent model we use only 1 recurrent layer. Furthermore, we split the training routine into two sections of equal number of epochs. In the first part, only the internal parameters of the network are optimized ($W_{in}$, $J$). In the second part, we train only the readout layer using cross-entropy loss. We also highlight that, according to our results, modifying $W_{back}$ during training results in severely degraded performance.

As for the multilayer perceptron, the structure of the model is the following:
$$
\text{Input} \to \text{Linear}(N) \to \text{sgn}(\tanh(\cdot)) \to \text{Linear}(N) \to \text{sgn}(\tanh(\cdot)) \to \text{Linear}(C).
$$
As stated in the main text, gradients completely ignore the sign function during the backward pass. This allows the model to be trained with regular backpropagation, despite having discrete activations.

For all experiments, optimal hyperparameter values were determined via a grid search, maximizing validation accuracy on a randomly selected holdout set of size \(P = 6000\).
The grid search was conducted for the lowest model width \(N = 1000\), and optimality was tested at other widths by applying small perturbations. 

We found that most hyper-parameters remained optimal across model sizes, except of $\lambda_y$, that was tuned accordingly.
\paragraph{\textbf{MNIST}} Images from the MNIST and FashionMNIST datasets are flattened to 1d arrays of size $D=784$. No other transformation is applied to the pixel values. The $P=60000$ images are equally split among $C=10$ classes. Optimal hyper-parameters are reported in Table \ref{tab:mnist}.
\begin{table}[h!]
    \centering
    \begin{tabular}{l|ll}
        \textbf{MNIST} & Binary MLP  & Ours \\
        \hline 
     Layer widths & 109,214,319,422,525,626 & 1000,2000,3000,4000,5000,6000 \\
     Hyperparameter values & $\eta=0.003$ & $\eta_J=0.058, \eta_{W_{in}}=0.159, \eta_{W_{back}}=0.0, \eta_{W_{out}}=0.005$ \\
                           & $w=0.0$ & $w_J=0.0, w_{W_{in}}=0.01$ \\
                           & $n_{epochs}=20$ & $n_{epochs}=20$ \\
                           &  & $J_D=0.95$ \\
                           &  & $\kappa_{W_{in}}=1.78, \kappa_{J}=1.78$ \\
                           &  & $\lambda_x=5.0$ \\
                           &  & $\lambda_y=\{1.62,1.3,1.1,0.8,0.8,0.8\}$ \\
    \end{tabular}
    \caption{Hyper-parameters used for experiments on the MNIST dataset. $w$ represents the weight decay coefficient, $\eta$ represent learning rates, $\kappa$ indicate Hebbian rule thresholds, while $\lambda$ represent external field strengths. Except for $\lambda_y$, whose optimal varies across layer widths, the other hyper-parameters are kept fixed.}
    \label{tab:mnist}
\end{table}

\paragraph{\textbf{FashionMNIST}} Images from the FashionMNIST dataset are treated exactly as the ones in the MNIST dataset. Optimal hyper-parameters are reported in Table \ref{tab:fmnist}.
\begin{table}[h!]
    \centering
    \begin{tabular}{l|ll}
        \textbf{FashionMNIST} & Binary MLP  & Ours \\
        \hline 
     Layer widths & 109,214,319,422,525,626 & 1000,2000,3000,4000,5000,6000 \\
     Hyperparameter values & $\eta=0.003$ & $\eta_J=0.118, \eta_{W_{in}}=0.17, \eta_{W_{back}}=0.0, \eta_{W_{out}}=0.0015$ \\
                           & $w=0.0$ & $w_J=0.0001, w_{W_{in}}=0.0002$ \\
                           & $n_{epochs}=20$ & $n_{epochs}=20$ \\
                           &  & $J_D=0.98$ \\
                           &  & $\kappa_{W_{in}}=1.74, \kappa_{J}=1.74$ \\
                           &  & $\lambda_x=4.91$ \\
                           &  & $\lambda_y=\{1.8,1.8,1.3,1.0,1.0,1.0\}$ \\
    \end{tabular}
    \caption{Hyper-parameters used for experiments on the FashionMNIST dataset. $w$ represents the weight decay coefficient, $\eta$ represent learning rates, $\kappa$ indicate Hebbian rule thresholds, while $\lambda$ represent external field strengths. Except for $\lambda_y$, whose optimal varies across layer widths, the other hyper-parameters are kept fixed.}
    \label{tab:fmnist}
\end{table}

\paragraph{\textbf{Entangled MNIST}}
We provide a description of the Entangled MNIST dataset, a more challenging compressed version of the MNIST dataset. 
The dataset is obtained by flattening the original MNIST images into vectors of dimension 784. Each vector is then projected onto a 100-dimensional subspace using a (fixed) random linear projection and later binarized to $\pm 1$ element-wise. The final dataset consists of $P=60000$ training instances and $P_{\text{eval}}=10000$ validation instances of dimension $D=100$, equally divided between $C=10$ classes. This process creates a more challenging learning scenario compared to the original MNIST, that tests the robustness of models operating on compressed and structurally entangled data, on which linear classifiers have poor performance.
As above, we provide optimal values for each hyperparameter in Table \ref{tab:emnist}.
\begin{table}[h!]
    \centering
    \begin{tabular}{l|ll}
        \textbf{Entangled-MNIST} & Binary MLP  & Ours \\
        \hline 
     Layer widths & 126,233,336,438,540,640 & 1000,2000,3000,4000,5000,6000 \\
     Hyperparameter values & $\eta=0.005$ & $\eta_J=0.02, \eta_{W_{in}}=0.3, \eta_{W_{back}}=0.0, \eta_{W_{out}}=0.005$ \\
                           & $w=0.0$ & $w_J=0.0, w_{W_{in}}=0.0$ \\
                           & $n_{epochs}=20$ & $n_{epochs}=20$ \\
                           &  & $J_D=0.90$ \\
                           &  & $\kappa_{W_{in}}=1.6, \kappa_{J}=1.4$ \\
                           &  & $\lambda_x=5.0$ \\
                           &  & $\lambda_y=\{3.5,2.5,1.75,2.0,1.75,1.25\}$ \\
    \end{tabular}
    \caption{Hyper-parameters used for experiments on the Entangled-MNIST dataset. $w$ represents the weight decay coefficient, $\eta$ represent learning rates, $\kappa$ indicate Hebbian rule thresholds, while $\lambda$ represent external field strengths. Except for $\lambda_y$, whose optimal varies across layer widths, the other hyper-parameters are kept fixed.}
    \label{tab:emnist}
\end{table}

\paragraph{\textbf{Tiny Imagenet features}} This dataset is obtained from the original Tiny imagenet subset of the ImageNet challenge \cite{deng2009imagenet}. Specifically, we extract feature vectors from the images with a fine-tuned convolutional backbone. The resulting data is composed of $P=100000$ one-dimensional vectors of dimension $D=512$, equally split among $C=200$ classes. As for the other dataset, we provide optimal hyper-parameters values in Table \ref{tab:timg}.
For the convolutional backbone, we choose a ResNet18 model \cite{resnet} pretrained on the full ImageNet Dataset and we fine-tune it on the Tiny Imagenet subset.
During training we apply the following data augmentation operations: Random Crop (224*224), random horizontal flipping ($p=0.50$), color jitting (brightness=0.4, constrast=0.4, saturation=0.4, hue=0.1), random grayscale ($p=0.1$), RandAugment \cite{randaugment} (num operations=2, magnitude=9), RandomErasing (p=0.25, scale=(0.02, 0.2), ratio=(0.3, 3.3)), CutMix ($p=0.5$) \cite{cutmix-paper} Mixup ($p=0.5$) \cite{mixup-paper}.
We train the model with stochastic gradient descent ($\eta=0.01, w=5\cdot10^{-4}$, momentum=0.9) for 50 epochs using a batch size of 64. The final model has a test accuracy of $0.765$.
We then extract the output of the Global Average Pooling at the end of the convolutional backbone with no data augmentation for the whole training and test set.
We do not apply any further operation.
The resulting feature vectors (and labels) are freely available at \url{https://huggingface.co/datasets/willinki/tinyimagenet-features}.

\begin{table}[h!]
    \centering
    \begin{tabular}{l|ll}
        \textbf{Tiny-Imagenet Features} & Binary MLP  & Ours \\
        \hline 
     Layer widths & 300,400,500,600,700,800 & 1155,1660,2216,2817,3458,4134 \\
     Hyperparameter values & $\eta=0.005$ & $\eta_J=0.02, \eta_{W_{in}}=0.3, \eta_{W_{back}}=0.0, \eta_{W_{out}}=0.005$ \\
                           & $w=0.0$ & $w_J=0.01, w_{W_{in}}=0.0$ \\
                           & $n_{epochs}=20$ & $n_{epochs}=20$ \\
                           &  & $J_D=0.90$ \\
                           &  & $\kappa_{W_{in}}=1.6, \kappa_{J}=1.4$ \\
                           &  & $\lambda_x=5.0$ \\
                           &  & $\lambda_y=\{3.1,2.5,2.5,2.25,2.25,2.25\}$ \\
                           &  & $\rho_J=0.99, \rho_{W_{in}}=0.99$ \\
    \end{tabular}
    \caption{Hyper-parameters used for experiments on the Tiny-Imagenet features dataset. $w$ represents the weight decay coefficient, $\eta$ represent learning rates, $\kappa$ indicate Hebbian rule thresholds, while $\lambda$ represent external field strengths. Except for $\lambda_y$, whose optimal varies across layer widths, the other hyper-parameters are kept fixed.}
    \label{tab:timg}
\end{table}

\subsubsection{Feature learning}
\begin{figure}
    \centering
    \includegraphics[width=0.5\linewidth]{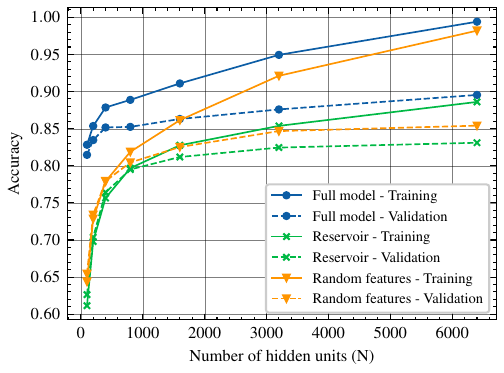}
    \caption{Training and validation accuracy against feature dimensionality for our proposed approach (blue), a random feature perceptron (orange) and a reservoir computing model (green).}
    \label{fig:old-bench}
\end{figure}

We assess more closely the ability of the proposed algorithm to perform feature learning. We compare its performance for the core module with two simple baselines that train a linear readout on fixed features. Specifically, these two baselines are:
\begin{itemize}
    \item \textbf{Perceptron}: composed of a random projection from the input dimension $D$ to the feature dimension $N$, followed by a element-wise sign function. A simple multi-class, one-versus-all perceptron is trained on these features.
    \item \textbf{Reservoir computing}: composed of a random projection $W^{in}$ from the input space $D$ to the feature space of size $N$. The model is then characterized by the following dynamics, defined by the random internal coupling matrix $J$, run until convergence:
    \[
        s_{i} = \text{sign}\left(\sum^{N}_{j \neq i} J_{ij} s_j +\sum^{N}_{j} W^{in}_{ij} x_j \right).
    \]
    A linear readout is trained on the fixed points of the dynamics. No update rules are in place for $W^{in}$ or $J$.
\end{itemize}
We compare training and test accuracy of these approaches while varying the feature dimension $N$ on the Entangled MNIST dataset. Results are reported in Figure \ref{fig:old-bench}.

We list here the hyperparameters used for the benchmarking experiments reported in the main text. 

A notable observation concerns the role of \(\lambda_x\). Comparable performance was achievable across a broad range \(\lambda_x \in [2.0, 6.0]\) when \(J_D\) and \(\lambda_y\) were adjusted accordingly. We selected \(\lambda_x = 5\) for the reported results, as it provided higher robustness to variations in the other two hyperparameters.

For the Random Features and Reservoir baselines, all hyperparameters remained optimal across all tested widths. We implement the baselines as special cases of the core module. Specifically, the reservoir model is obtained from the core module by setting $\eta_{W_{in}}=\eta_{J}=\lambda_y=0$. The random features model is obtained by taking the reservoir model and setting the maximum amount of dynamics iterations $\text{max}_{\text{iter}} = 1$.

The final chosen parameters are summarized in Table~\ref{tab:hyperparameters}.
\begin{table}
    \centering
    \begin{tabular}{l|llll}
            \hline
            Hyperparameter & $N=\{100,\dots,3200\}$ & $N=6400$ & Random Features & Reservoir\\
            \hline
            $J_D$ & 0.5 & 0.5 & / & 0.5 \\
            $\eta_{J}$ & 0.005 & 0.01 & 0.0 & 0.0 \\
            $\eta_{W_{in}}$ & 0.03 & 0.01 & 0.0 & 0.0  \\
            $\eta_{W_{out}}$ & 0.03 & 0.01 & 0.03 & 0.03 \\
            $\kappa_{J}$ & 1.4 & 1.8 & / & / \\
            $\kappa_{W_{in}}$ & 3.0 & 3.0 & 3.0 & 3.0 \\
            $\kappa_{W_{out}}$ & 3.0 & 3.0 & 3.0 & 3.0 \\
            $\lambda_{x}$ & 5.0 & 5.0 & 5.0 & 5.0 \\
            $\lambda_{y}$ & 0.9 & 0.7 & 0 & 0 \\
            $n_{epochs}$ & 200 & 400 & 200 & 200 \\
            $\text{max}_{iter}$ & 5 & 5 & 1 & 5 \\
            \hline
        \end{tabular}
    \caption{Optimal hyperparameters for models used in benchmarking. Both random features and reservoir models can be seen as a special case of the full model with some specific components removed.}
    \label{tab:hyperparameters}
\end{table}

\subsubsection{Hetero-association task}
To evaluate the hierarchical processing capabilities of the model, we study its capacity to perform hetero-associative memory tasks. Given a dataset \( \{ x^\mu, y^\mu \}_{\mu=1}^{P} \), the hetero-association task consists in learning to map each \(x^\mu\) to the corresponding \(y^\mu\). We define the \emph{hetero-association capacity} \( P_{\mathrm{max}} \) as the largest dataset size for which the network reliably maps random input-output associations.

Consider a network consisting of \( L \) layers, each composed of \( N \) hidden units, and a dataset \( \{ x^\mu, y^\mu \}_{\mu=1}^{P} \), where each \( x^\mu \in \{-1, 1\}^{N} \) and \( y^\mu \in \{-1, 1\}^{N} \) is independently sampled from a Rademacher distribution. 
The network is trained according to the procedure detailed in the main text. It produces outputs \( \hat{s}^{*\mu} \) given inputs \( x^\mu \), . 
For fixed \( N \), we define \( P_{\mathrm{max}} \) as the largest value of \( P \) for which at least 95\% of the patterns satisfy the alignment criterion:
\begin{equation}
    \hat{q}^\mu := \frac{1}{N} \sum_{i=1}^{N} \hat{s}^{*\mu}_i y^\mu_i > 1 - \varepsilon,
\end{equation}
with \( \varepsilon = 0.05 \) fixed throughout.

\begin{figure*}[ht!]
    \centering
    \includegraphics[width=0.50\linewidth]{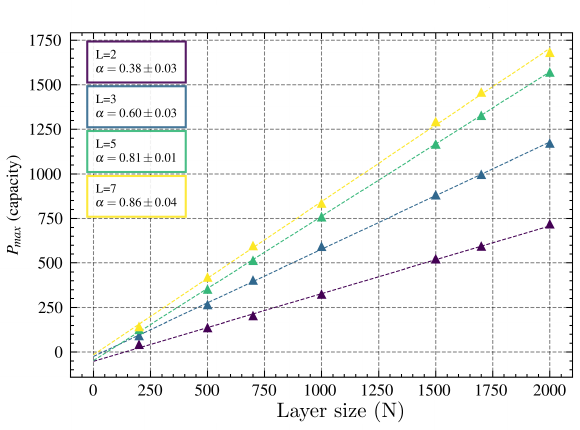} \,\,\,
    \includegraphics[width=0.42\linewidth]{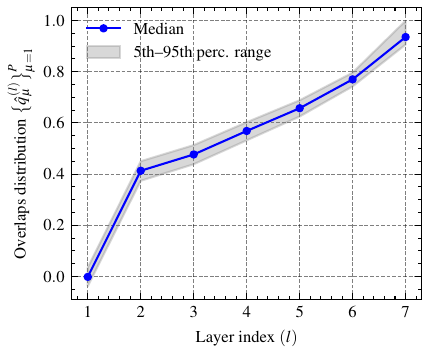}
    \caption{\textbf{(a)} Hetero-association capacity: Maximum dataset size $P_{\mathrm{max}}$ for successful hetero-association as a function of the number of hidden units $N$, across different network depths $L$. Each line corresponds to a different number of layers $L$, and the slope $\alpha$ of the linear fit quantifies capacity scaling. Performance is measured under the criterion $\frac{1}{N} \sum_{i=1}^{N} \hat{s}^{*\mu}_i y^\mu_i > 1-\varepsilon$ for at least 95\% of patterns, with $\varepsilon = 0.05$. \textbf{(b)} Hierarchical overlaps: Overlap between internal network states and output patterns across network layers. Data were obtained by selecting the 10 trials closest to $P_{\mathrm{max}}$ for networks with $L=7$ and $N=2000$. The median overlap $\hat{q}_\mu^{(l)}$ across patterns (blue curve) is shown, together with the shaded region indicating the 5th–95th percentile range. The internal state at layer $l=1$ is closely aligned with the input and de-correlated from the output, while deeper layers progressively align more strongly with the output, clearly demonstrating the hierarchical processing capabilities of the trained model.}
    \label{fig:combined}
\end{figure*}

To estimate \( P_{\mathrm{max}} \) empirically, we train the model repeatedly with dataset sizes \( P \) sampled uniformly from the interval \([N/2, N]\), and compute the distribution of overlaps $\hat{q}^\mu$ over the patterns. For each trial, we extract the 5th percentile of the overlap distribution $\mathcal{O}_5(P)$.
We then fit a smooth spline to the curve \( \mathcal{O}_5(P) \) and estimate \( P_{\mathrm{max}} \) as the value of \( P \) for which \( \mathcal{O}_5(P) = 1 - \varepsilon \).
Finally, to quantify how capacity scales with the number of hidden units, we fit a linear model to the relationship \( P_{\mathrm{max}}(H) \), and extract the scaling coefficient for each network depth \( L \).
Results are reported in Figure~\ref{fig:combined}(a).

To illustrate the hierarchical processing capabilities of our model, we selected the 10 trials closest to $P_{\mathrm{max}}$ for networks with $L=7$ and $N=2000$. 
We computed the overlap between the internal states of each layer and output patterns, averaging these overlaps across the chosen trials. 
Specifically, we define the overlap $\hat{q}_\mu^{(l)}$ between the internal state at layer $l$ for pattern $\mu$ and the corresponding output $y^\mu$ as:
\begin{equation}
\hat{q}_\mu^{(l)} = \frac{1}{N} \sum_{i=1}^{N} \hat{s}_i^{*(l),\mu} y_i^\mu\,,
\end{equation}
where $\hat{s}_i^{*(l),\mu}$ denotes the final state of the $i$-th hidden unit at layer $l$ for pattern $\mu$, and $y_i^\mu$ denotes the $i$-th component of the target vector associated with pattern $\mu$. The resulting averaged overlaps clearly show that the internal state in the first layer remains closely aligned with the input and essentially de-correlated from the output, while deeper layers systematically approach and ultimately align closely with the target output (see Fig.~\ref{fig:combined}(b)).

%% file: importance-stable-manifold.tex
\label{sec:symmetry-irrelevant}
In this section, we report on the second set of experiments concerning the role of the RM geometry and accessiblity.
We consider the classification task on Entangled MNIST, and we consider several variants of our proposed algorithm. All variants are based on the core model and the learning algorithm discussed in the main text. First, we consider two variants that differ in the initialization of the internal couplings:
\begin{itemize}
    \item \textbf{asym}: this is the core module with $\rho = 1$; the internal coupling matrix is initialized sampling each off-diagonal component i.i.d. from a Gaussian, \(J_{ij} \sim \mathcal{N} \left(0, \frac{1}{\sqrt{N}} \right) \), and diagonal entries are equal to \(J_D\).
    \item \textbf{sym0}: the coupling matrix is initialized symmetrically: sample \(J\) like for the \textbf{asym} baseline, compute \(J_{sym} = (J+J^T)/\sqrt{2}\), and set diagonal entries of \(J_{sym}\) to \(J_D\).
\end{itemize}
Additionally, we consider two more variants that use the same initialization as in \textbf{sym0}, but additionally implement a variant of the update rule \eqref{eq:plasticity_rule} that preserves the initial symmetry:
\begin{itemize}
    \item \textbf{sym1}: Condition the update of \(J_{ij}\) on both \(s_i\) and \(s_j\) not being stable enough, instead of \(s_i\) only. Training with batches of examples, this is done independently for each example in the batch. In formulas: \[ J_{ij} \leftarrow J_{ij} + \, \eta \, s_i^{*} \, s_j^{*} \; \mathds{1}(s_i^{*} \cdot h_i \le k \And s_j^{*} \cdot h_j \le k) \] Note that we have omitted \(A_{ij}\) since the core model is fully connected.
    \item \textbf{sym2}: Update \(J_{ij}\) using the average of the updates that \eqref{eq:plasticity_rule} would have made for \(J_{ij}\) and \(J_{ji}\). In formulas: \[ J_{ij} \leftarrow J_{ij} + \frac12 \, \eta \, s_i^{*} \, s_j^{*} \; \mathds{1}(s_i^{*} \cdot h_i \le k) + \frac12 \, \eta \, s_i^{*} \, s_j^{*} \; \mathds{1}(s_j^{*} \cdot h_j \le k)\]
\end{itemize}

\begin{figure}[h!]
  \centering
  \begin{subfigure}{0.49\linewidth}
    \centering
    \includegraphics[width=\linewidth]{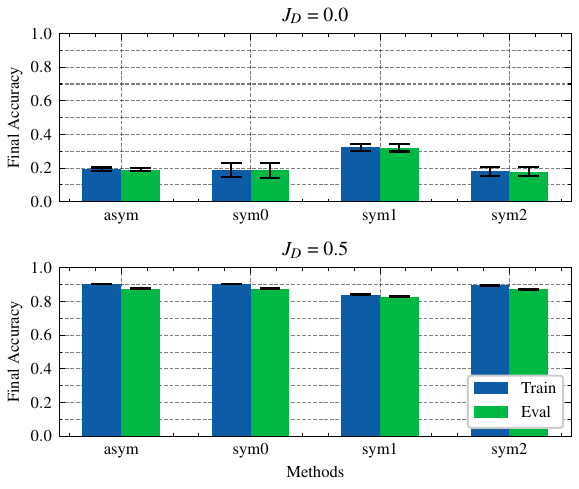}
    \label{fig:symmetric-rules-main}
  \end{subfigure}\hfill
  \begin{subfigure}{0.49\linewidth}
    \centering
    \includegraphics[width=\linewidth]{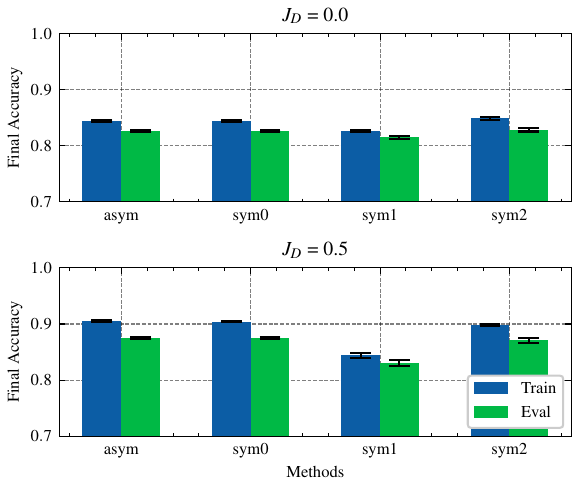}
    \label{fig:symmetric-rules-app}
  \end{subfigure}
  \caption{Impact of the existence of the accessible stable region vs symmetry of the internal couplings on performance on Entangled MNIST for the core model. \textbf{Left}: Algorithm of the main text (transient supervision). \textbf{Right}: Approximate training scheme described in the Supplementary Material}
  \label{fig:symmetric-rules-both}
\end{figure}

We consider \(J_D = 0.0, 0.5\) and we compare the average performance on Entangled MNIST of the methods across 3 seeds. The results are shown in Figure \ref{fig:symmetric-rules-both}. The key observation is that symmetry has little-to-no impact on performance, despite the dynamics being convergent from the beginning of training; on the other hand, introducing a self-coupling, thereby leading to the onset of the RM, produces a drastic improvement in performance for the asymmetric baseline and the symmetric variants alike.

%% file: evolution-symmetricity.tex
\label{sec:evolution-symmetricity}
In this section, we describe how the degree of symmetry of the internal-couplings matrix evolves as training progresses, with symmetric or asymmetric initialization. The models considered thus correspond to those called \textbf{asym} and \textbf{sym0} in the previous section.

\begin{figure}[h!]
    \centering
    \includegraphics[width=1.0\linewidth]{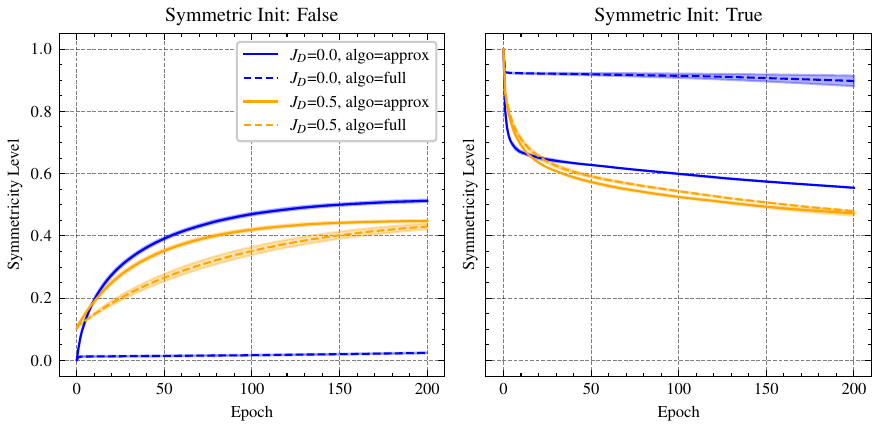}
    \caption{Evolution of the symmetry level \eqref{eq:sym-level} of the matrix of internal couplings during training of the core module. We consider $J_D = 0.0, 0.5$, and both the full (main text) and approximate (SM) algorithms.
    \textbf{Left}: Asymmetric initialization for $J$ ($\rho = 1$). \textbf{Right}: Symmetric initialization for $J$ (sym0 in previous section).
    }
    \captionsetup{format=plain, font=small, labelfont=bf}
    \label{fig:sym-evolution-both}
\end{figure}

To quantify the symmetry level of a square matrix of internal couplings \(J\), we use the following metric. We decompose \(J\) into its symmetric and anti-symmetric part, according to \[J = S + A = (J + J^T)/2 + (J - J^T) / 2\] It can be shown that \(S\) so defined is the best symmetric approximation of \(J\) in the Frobenius sense. Then, we use 
\begin{equation}
    \label{eq:sym-level}
    \rho = (||S||-||A||) / (||S|| + ||A||)
\end{equation}
as a measure of the symmetry level of \(J\). \(\rho\) is normalized between \(-1\) and \(1\); it equals 1 when the matrix is perfectly symmetric and -1 when it is perfectly anti-symmetric. A matrix with random iid entries from a Gaussian has a score close to 0. We show the evolution of the symmetry level during training in Figure \ref{fig:sym-evolution-both}.

%% file: landscape_stability.tex
\label{sec:landscape-analysis}

The local entropy analysis predicts that the  models  exhibit a RM for sufficiently large $J_D$ or $\lambda$.  
We study how the training process affects this landscape by analyzing a finite-size network ($N=1600$, $L=1$) trained on the Entangled-MNIST dataset with the hyperparameters reported in Table~\ref{tab:hyp-landscape}.

\begin{table}[h]
    \centering
    \begin{tabular}{l|cccccccccccc}
    \hline
    \textbf{Hyperparameter}  & $N$ & $L$ & $J_D$ & $\eta_{J}$ & $\eta_{W_{in}}$ & $\eta_{W_{out}}$ & $\kappa_{J}$ & $\kappa_{W_{in}}$ & $\kappa_{W_{out}}$ & $\lambda_{x}$ & $\lambda_{y}$ & $n_{\text{epochs}}$ \\
     \textbf{Value} & 1600 & 1 & 0.5 & 0.005 & 0.01 & 0.01 & 0.9 + $J_D$ & 3.0 & 3.0 & 2.0 & 0.9 & 30 \\
    \hline
    \end{tabular}    
\caption{Hyperparameters for the Entangled-MNIST landscape analysis. The trained network achieved $88.54\%$ training accuracy and $85.92\%$ validation accuracy.}
    \label{tab:hyp-landscape}
\end{table}

We select a random subset of $P=2000$ training inputs $\{x^\mu\}_{\mu=1}^P$.  
For each $x^\mu$, we compute the corresponding stable state $\hat{s}^{*\mu}$ by iterating the dynamics update rule
\begin{equation}
    s_i \leftarrow \mathrm{sign}\!\left(\sum_{j\neq i} J_{ij} s_j + J_D s_i + \lambda_x x^\mu_i\right)
    \label{eq:dynamics}
\end{equation}
until convergence.

To assess the connectivity between two stable states $(\hat{s}^{*\mu}, \hat{s}^{*\nu})$, we construct intermediate input configurations $x^{\mu\nu}_\tau$ with $\tau \in [0,1]$ drawn uniformly from
\begin{equation}
    \left\{x \in \{-1,+1\}^N \; \middle| \; \frac{d(x,x^\mu)}{d(x^\mu,x^\nu)} = \tau \right\},
    \label{eq:intermediate}
\end{equation}
where \(d(\cdot,\cdot)\) is the Hamming distance.  
Evolving $x^{\mu\nu}_\tau$ under (\ref{eq:dynamics}) gives the intermediate fixed point \(\hat{s}^{*\mu\nu}_\tau\).

We quantify deviations from a geodesic path in state space through the \emph{normalized path curvature}
\begin{equation}
    C^{\mu\nu}(\tau) =
    \frac{d(\hat{s}^{*\mu},\hat{s}^{*\mu\nu}_\tau) + d(\hat{s}^{*\mu\nu}_\tau, \hat{s}^{*\nu})}
         {d(\hat{s}^{*\mu},\hat{s}^{*\nu})}.
\end{equation}
For each pair, we take the maximum along the path,
\begin{equation}
    \kappa^{\mu\nu} = \max_{\tau\in(0,1)} \mathbb{E}[C^{\mu\nu}(\tau)],
\end{equation}
where $\mathbb{E}[\cdot]$ is the empirical average over resamplings of $x^{\mu\nu}_\tau$.

\begin{figure}[t]
  \centering
  \begin{minipage}[t]{0.48\linewidth}
    \centering
    \includegraphics[width=\linewidth]{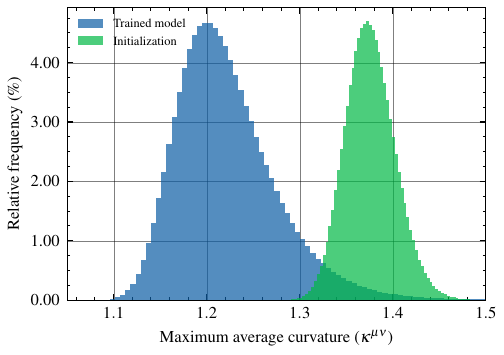}
  \end{minipage}\hfill
  \begin{minipage}[t]{0.48\linewidth}
    \centering
    \includegraphics[width=\linewidth]{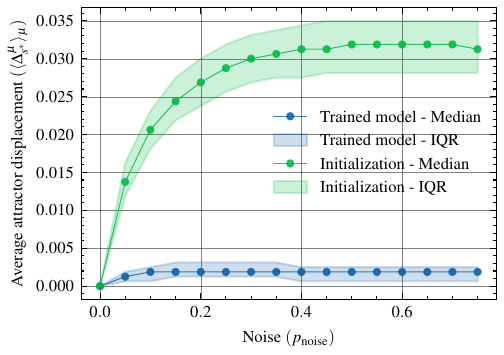}
  \end{minipage}
  \caption{(a) Distribution of the maximum normalized geodesic curvature $\kappa^{\mu\nu}$ between stable attractors, before (green) and after (blue) training. Training reduces typical curvature, indicating straighter connecting paths. (b) Distribution of the attractor displacement $\Delta^\mu_{s^*}$ as a function of the noise injected into $\hat{s}^{*\mu}$, showing increased robustness post-training. The median is plotted together with the 25th-75th percentile range (IQR).}
  \label{fig:landscape-curvature-hist}
\end{figure}

\begin{figure}[t]
  \centering
  \begin{minipage}[t]{0.80\linewidth}
    \centering
    \includegraphics[width=\linewidth]{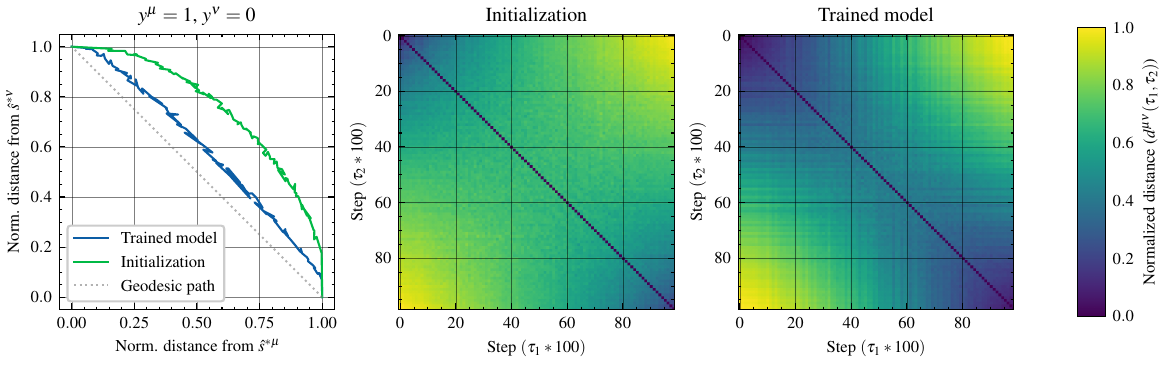}
  \end{minipage}\hfill
  \begin{minipage}[t]{0.80\linewidth}
    \centering
    \includegraphics[width=\linewidth]{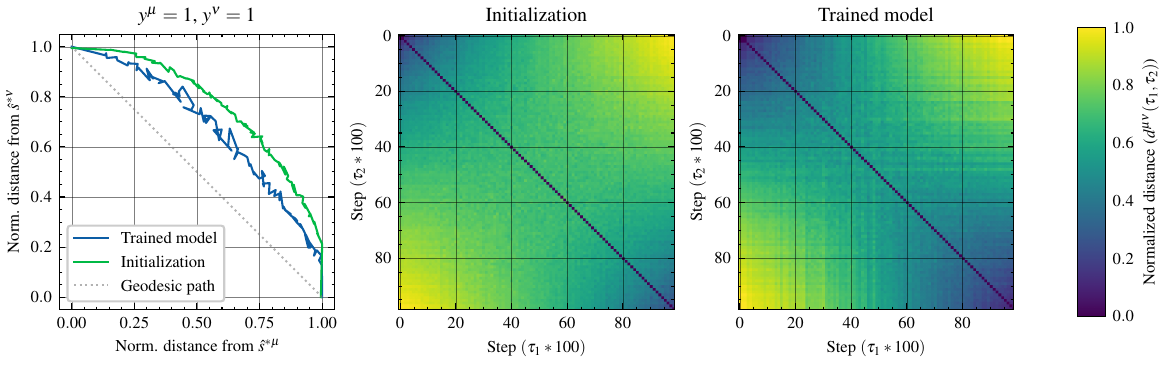}
  \end{minipage}
  \begin{minipage}[t]{0.80\linewidth}
    \centering
    \includegraphics[width=\linewidth]{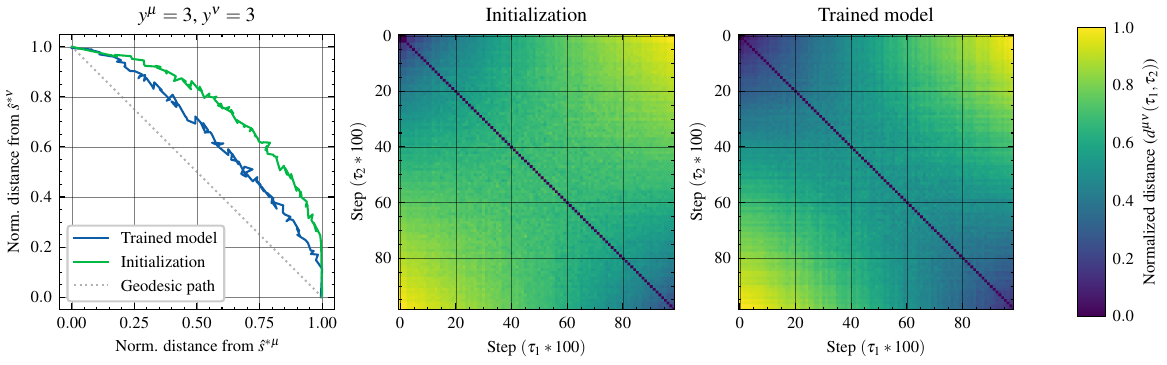}
  \end{minipage}
  \begin{minipage}[t]{0.80\linewidth}
    \centering
    \includegraphics[width=\linewidth]{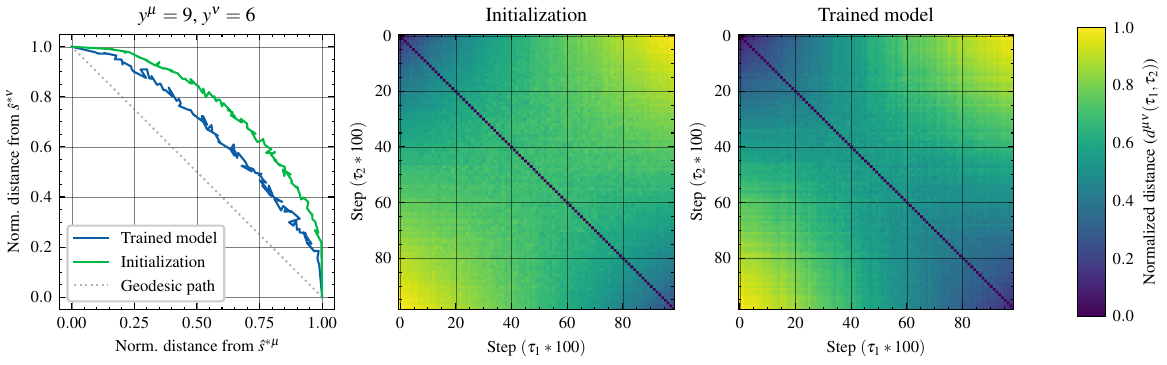}
  \end{minipage}
  \caption{Pairwise normalized distances $d^{\mu\nu}(\tau_1,\tau_2)$ along interpolated trajectories between two training examples $(x^\mu, y^\mu), (x^\nu, y^\nu)$. After training, trajectories show shorter paths, closer to the actual geodesic.}
  \label{fig:trajectories}
\end{figure}

Figure~\ref{fig:landscape-curvature-hist}(a) shows a significant distribution shift of $\kappa^{\mu\nu}$ after training:  
from mean $1.367$ (s.d.\ $0.027$) at initialization to mean $1.222$ (s.d.\ $0.055$) post-training.  
Values close to one imply that intermediate attractors lie near the shortest Hamming path between endpoints, suggesting that, as a result of training, attractors associated with different inputs are connected by low-curvature corridors.

To characterize these connections more directly, we compute the \emph{normalized trajectory distance}
\begin{equation}
D^{\mu\nu}(\tau_1,\tau_2) =
\frac{d(\hat{s}^{*\mu\nu}_{\tau_1}, \hat{s}^{*\mu\nu}_{\tau_2})}
     {d(\hat{s}^{*\mu}, \hat{s}^{*\nu})}.
\end{equation}
Representative cases in Fig.~\ref{fig:trajectories} show that, after training, the trajectories between attractors exhibit reduced length and curvature, in agreement with the global curvature statistics reported in Figure \ref{fig:landscape-curvature-hist}(a).

We also assess local robustness by perturbing each stable state with spin-flip noise:
\[
s^\mu_{\text{noise}} = \hat{s}^{*\mu} \odot \xi_{p_{\text{noise}}}, 
\quad \xi_i \sim \mathrm{Rademacher}(p_{\text{noise}}),
\]
where \(p_{\text{noise}}\) is the flip probability.  
From $s^\mu_{\text{noise}}$, we evolve the system with the update rule in Equation \ref{eq:dynamics} to its fixed point \(\tilde{s}^{*\mu}\) and measure the \emph{attractor displacement}
\[
\Delta^{\mu}_{s^*} = d(\hat{s}^{*\mu}, \tilde{s}^{*\mu}).
\]
The distribution of $\Delta^{\mu}_{\text{att}}$ [Fig.~\ref{fig:landscape-curvature-hist}(b)] shifts toward smaller values after training, indicating greater stability to perturbations.  
Even before training, displacements remain fairly small ($<5\%$), consistently with the presence of highly entropic clusters of stable attractors, whose basin of attraction is determined by the left field.

Taken together, the evidence suggests that training reorganizes the RM by shortening and straightening inter-attractor paths, enlarging the effective reach of individual attractors, and improving robustness to perturbations.

\begin{figure}
    \centering
    \begin{subfigure}[t]{0.63\linewidth}
        \vspace{0pt}
        \centering
        \includegraphics[width=\linewidth]{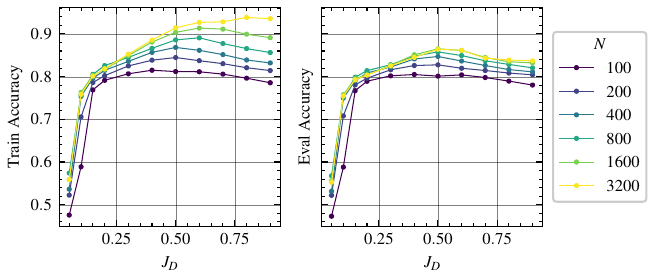}
        \label{fig:role-of-jd-plot}
    \end{subfigure}%
    \hspace{0.01\linewidth} 
    \begin{subfigure}[t]{0.35\linewidth}
        \vspace{2pt}
        \centering
        \begin{tabular}{ll}
            \hline
            Hyperparameter & Value \\
            \hline
            $\eta_{J}$ & 0.005 \\
            $\eta_{W_{in}}$ & 0.01 \\
            $\eta_{W_{out}}$ & 0.01 \\
            $\kappa_{J}$ & 0.9 + $J_D$ \\
            $\kappa_{W_{in}}$ & 3.0 \\
            $\kappa_{W_{out}}$ & 3.0 \\
            $\lambda_{x}$ & 5.0 \\
            $\lambda_{y}$ & 0.9 \\
            $n_{epochs}$ & 100 \\
            \hline
        \end{tabular}
    \end{subfigure}
    \caption{(a) Training and evaluation accuracy as a function of the self-coupling parameter $J_D$ for various hidden layer sizes $N \in \{100, 200, 400, 800, 1600, 3200\}$. The network shows optimal performance for intermediate values of $J_D$, with larger models benefiting more from self-coupling. (b) Hyperparameter configuration used across all experiments.}
    \label{fig:role-of-jd}
\end{figure}

\label{sec:role-of-jd}
Finally, we discuss the important effect of the self-coupling constant \(J_D\) on training performance. To do so, we train our recurrent neural network on the Entangled-MNIST dataset while varying the value of \(J_D\) in the interval \([0.05, 0.9]\). In accordance with the main text, we choose \(L=1\) and \(N \in \{100, 200, 400, 800, 1600, 3200\}\). We train all models using an identical set of hyperparameters, detailed in Figure~\ref{fig:role-of-jd}. Training and validation accuracies are reported in Figure~\ref{fig:role-of-jd}.

As expected, we observe optimal performance at intermediate values of \(J_D\). In the limit \(J_D \to 0\), the RM geometry collapses, and the model effectively regresses to random guessing. On the other hand, as \(J_D \to \infty\), the internal dynamics lose their discriminative capacity, since every state becomes a stable fixed point. In this regime, the network effectively behaves as a random feature model, as the internal state \(s\) at time \(t=1\) becomes fully determined by the input \(\xi^\mu\) (or more generally, by its projection \(W_{\text{in}} \xi^\mu\)).

Between these two extremes, however, lies a nontrivial intermediate regime where the RM remains exploitable and the internal dynamics play a meaningful role in shaping the model's performance. Additionally, we find that increasing the hidden layer size \(N\) enhances trainability and stabilizes training accuracy across a broader range of \(J_D\), although the accuracy gains are more pronounced with training patterns than it is for unseen ones.

Finally, we note that the optimal value of \(J_D\) depends on multiple factors, the most prominent being the input coupling strength \(\lambda_x\).

%% file: evolution-of-internal-states.tex
The learning process consists in an iterated stabilization of the network attractors, carried via the perceptron rule.
The network dynamics undergoes a first supervised phase, where a stable state $s^\prime$ is reached, followed by an unsupervised phase ($\lambda_y$=0), where the network reaches the final attractor $s^*$.
The sucess of this learning process is implicitly based on the observation that, when switching between supervised and unsupervised relaxation,  the stable configurations remain  almost unaltered. Only the local fields change.
In other words, in order to have an effective supervising signal, $s^\prime$ and $s^*$ must be highly correlated.

To verify directly this scenario,  we analyze the internal states of the network at different stages of training. 
Specifically, we study the \emph{overlap} between the network states \( s^\prime \) and \( s^* \), which correspond to the fixed points reached during the supervised and unsupervised phases, respectively (see Algorithm ~\ref{alg:supervised_learning}).

The average overlap on the dataset is defined as:
\[
    q_{\text{dyn}} = \frac{1}{NP}\sum_{\mu=1}^p \sum_{i=1}^N s_i^{\prime\mu} s_i^{*\mu},
\]
We evaluate the distribution of overlaps during learning using a single-layer network (\(L = 1\)) with \(N = 1600\) neurons, trained on the Entangled-MNIST dataset. 
We vary the strength of internal coupling (\(J_D\)), input field (\(\lambda_x\)), and right field (\(\lambda_y\)). 
For each setting, we measure the overlap distribution at initialization, after the first epoch, and at the end of training (epoch 20). 

Results are summarized in Table~\ref{tab:sprime_star}.
\begin{table}[ht]
\begin{tabular}{lllcccc}
    \hline
    $J_D$ & $\lambda_x$ & $\lambda_y$ & Initialization & Epoch 1 & Epoch 20 \\
    \hline
    0.5 & 5.0 & 0.9 & 
    $0.930~[0.921, 0.938]$ & 
    $0.991~[0.987, 0.995]$ & 
    $0.996~[0.993, 0.999]$ \\

    0.5 & 5.0 & 0.5 & 
    $0.975~[0.970, 0.980]$ & 
    $0.998~[0.996, 0.999]$ & 
    $0.999~[0.999, 1.000]$ \\

    0.5 & 2.0 & 0.9 & 
    $0.761~[0.744, 0.779]$ & 
    $0.995~[0.991, 0.998]$ & 
    $0.999~[0.996, 1.000]$ \\

    0.5 & 2.0 & 0.5 & 
    $0.892~[0.877, 0.907]$ & 
    $0.999~[0.998, 1.000]$ & 
    $1.000~[0.999, 1.000]$ \\

    0.0 & 5.0 & 0.9 & 
    $0.862~[0.852, 0.871]$ & 
    $1.000~[0.999, 1.000]$ & 
    $1.000~[1.000, 1.000]$ \\

    0.0 & 5.0 & 0.5 & 
    $0.915~[0.907, 0.922]$ & 
    $1.000~[1.000, 1.000]$ & 
    $1.000~[1.000, 1.000]$ \\

    0.0 & 2.0 & 0.9 & 
    $0.649~[0.634, 0.664]$ & 
    $1.000~[1.000, 1.000]$ & 
    $1.000~[1.000, 1.000]$ \\

    0.0 & 2.0 & 0.5 & 
    $0.748~[0.734, 0.761]$ & 
    $1.000~[1.000, 1.000]$ & 
    $1.000~[1.000, 1.000]$ \\
    \hline
\end{tabular}
\caption{Average overlap \( q_{\text{dyn}} \) between the network states \( s^\prime \) (supervised phase) and \( s^* \) (unsupervised phase) across training. 
Each entry reports the median overlap and its 5th-95th percentile range over the training set of Entangled-MNIST. 
We vary the self-coupling strength \( J_D \), left field \( \lambda_x \), and right field \( \lambda_y \), and track the evolution of the overlap at initialization, after the first training epoch, and after 20 epochs.}
\label{tab:sprime_star}
\end{table}

\renewcommand{\arraystretch}{1.0}
\setlength{\tabcolsep}{6pt}
Optimal training performance (training accuracy 91\%, validation accuracy 87\%) is obtained with $J_D=0.5$, $\lambda_x=5.0$ and $\lambda_y=0.9$ (first row). Any model with $J_D=0.0$ is unable to learn effectively (training/validation accuracy $\sim0.1$). 
The results show that the learning scheme improves the stability of fixed points already after the first epoch. Moreover, optimal learning occurs when the supervisory signal is strong enough to nudge the attractor without fully overriding the intrinsic dynamics.
Lastly, the results suggest that stabilizing attractor states alone is not sufficient for successful learning, emphasizing the importance of attractor geometry for the classification task. \\

In addition to the evolution of the network configuration during the relaxation dynamics, it is natural to wonder about the evolution of the internal representations associated with each input-output pair throughout training. To explore this aspect, we consider a random subset of \(P=300\) patterns from Entangled-MNIST and we measure the similarity (1 minus the normalized Hamming distance) between the internal representations associated to different patterns before and after training. To visualize the data, we order the patterns by label (first all zeros, then all ones, etc.) and we use a heatmap to encode the similarity matrix.

\begin{figure*}[ht!]
    \centering
    \includegraphics[width=\textwidth]{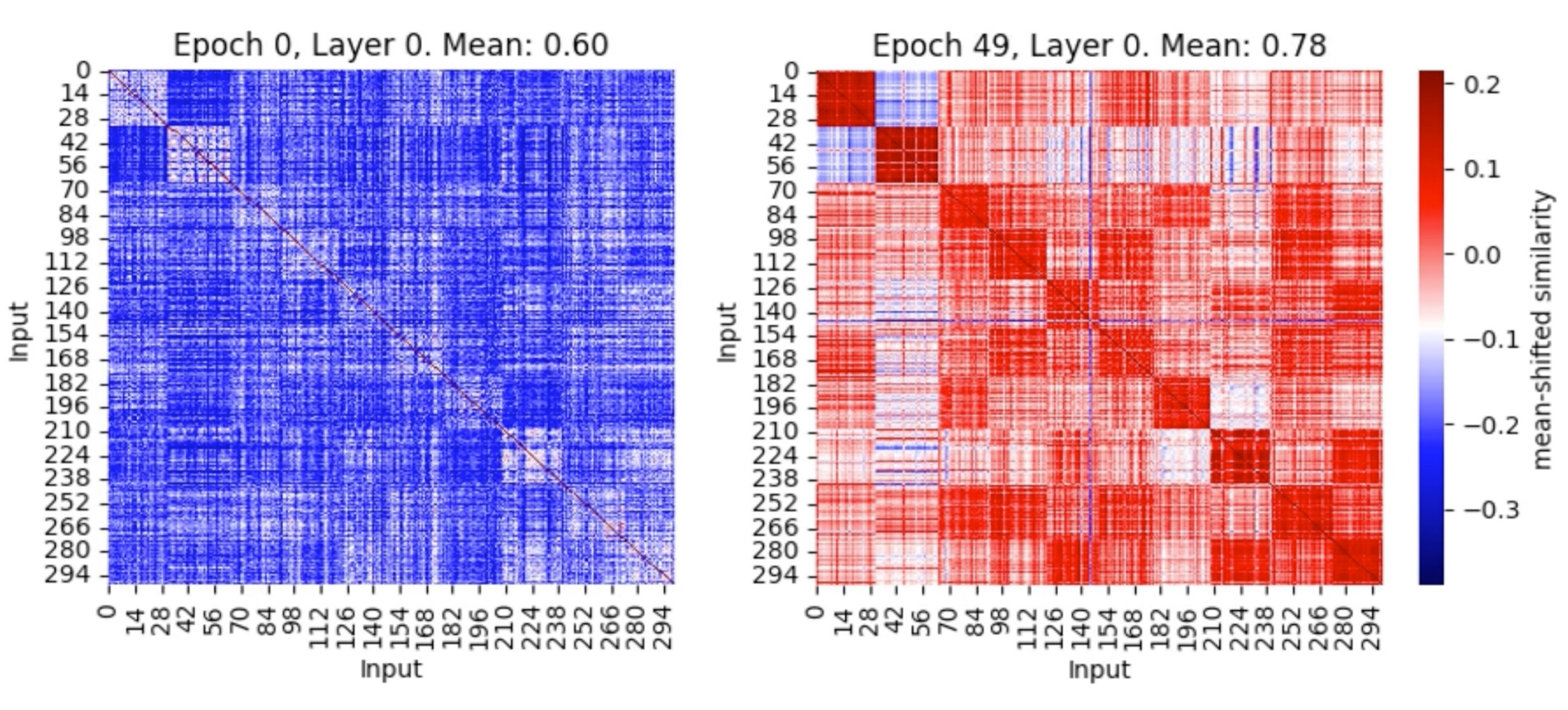}
    \caption{Similarity matrix of the internal representations of a fixed set of 300 patterns from Entangled MNIST before (\textbf{left}) and after (\textbf{right}) training. Patterns are ordered according to their label. The similarity measure used is one minus the hamming distance; the similarities are shifted by a single mean value, in such a way that colors remain consistent across time. The mean value reported on top of each heatmap is the average similarity across all pairs at that time, not shifted.}
    \label{fig:representations-heatmap}
\end{figure*}

More precisely, let \(x_\mu\) and \(x_\nu\) be two patterns with indexes \(\mu\) and \(\nu\) in the chosen ordering. We run inference with the model on \(x_\mu\), which consists in letting the network relax under the influence of \(x_\mu\) as left external field, and without any right external field, until convergence. We obtain an internal representation \(s^*_\mu\) for \(x_\mu\); we do the same for \(x_\nu\) to obtain \(s^*_\nu\). The entry \((\mu, \nu)\) of the similarity matrix is: \(S_{\mu \nu} = 1 - d(s^*_\mu, s^*_\nu)\). We compute the similarity matrix before and after training; for visualization purposes, we shift all values of all similarity matrices by a single mean value, computed averaging across pattern pairs and across time. The results are shown in Figure \ref{fig:representations-heatmap}. Two trends are observed, comparing representations before and after training: first, the representations of different patterns tend in general to get closer together, occupying a cone in state space; and second, the representations become strongly clustered based on label (evidenced by the checker structure of the matrix)with patterns sharing the same label being significantly more similar to each other than patterns having different labels.

